\documentclass[10pt]{article} 
\usepackage{a4,epsf,latexsym}

\begin{document}

\title{Bose-Einstein Condensates in Optical Lattices: 
	Spontaneous Emission in the Presence of Photonic Band Gaps.
}

\author{
 Karl-Peter Marzlin\\
   {\em {\small Fachbereich Physik der Universit\"at Konstanz, 
   Postfach 5560 M674, D-78457 Konstanz, Germany}}\\
 Weiping Zhang \\
   {\em {\small Centre for Lasers and Applications, and Department of Physics,
   }}\\
   {\em {\small Macquarie University, Sydney, New South Wales 2109, Australia}}
}
\maketitle

$ $ 
Shortened title: Spontaneous Emission in Bose-Einstein Condensates

\begin{abstract}
An extended Bose-Einstein condensate (BEC) in an optical lattice provides 
a kind of periodic dielectric and causes band gaps to occur in the spectrum
of light propagating through it. We examine the question whether
these band gaps can modify the spontaneous emission rate of atoms excited
from the BEC, and whether they can lead to a self-stabilization of the 
BEC against spontaneous emission. 
We find that self-stabilization is not possible for BECs with a density
in the order of $10^{14}$ cm$^{-3}$. However, the corresponding 
non-Markovian behavior produces significant effects in the decay of 
excited atoms even for a 
homogeneous BEC interacting with a weak laser beam. These
effects are caused by the occurrence of an avoided
crossing in the photon (or rather polariton) spectrum. 
We also predict a new channel for spontaneous decay which arises from 
an interference between periodically excited atoms and periodic photon modes.
This new channel should also occur in ordinary periodic dielectrics.
\end{abstract}

\section{Introduction}
It is well-known that the radiation properties of atoms can dramatically
be manipulated by changing the environment where atoms emit photons.
For micro-cavities it has been demonstrated \cite{cavity} and for
periodic dielectric media predicted \cite{john94,kofman94}
that a suppression of spontaneous emission (SE) can be achieved.
In the case of a micro-cavity this happens because its geometry reduces
the radiation-mode density, whereas in a periodic dielectric medium
SE is suppressed due to the formation of photonic band gaps (PBG).

The recent achievement of Bose-Einstein condensation in magnetic traps
\cite{experimente} has provided a new state of matter where all atoms
share a single macroscopic quantum state. Such a state of matter offers
great opportunities to explore and test new phenomena related to macroscopic
quantum coherence. Recently several authors have theoretically studied
spontaneous emission in a trapped BEC.
In this case the continuous center-of-mass 
momentum distribution leads to an increase of SE 
\cite{javanainen93,cooper95}. In addition, the stimulated emission can
be increased by the Bose enhancement in a BEC \cite{hope97}.
In the case of two BECs, interference effects can be important
\cite{savage97}.

The present work is focused on the case of an extended BEC and
was motivated by the following idea. If an extended
BEC is placed
in an optical lattice it will become periodic. Since such a BEC does provide
a (quantum) dielectric it affects the properties of photons
propagating through it and phenomena similar to PBGs do occur 
\cite{marzlin98b}. Since then the
photon mode density around the resonance frequency is reduced
one can expect that SE is suppressed by non-Markovian effects. 
We thus are led to the following 
question: can a BEC in an optical lattice stabilize itself against
spontaneous emission?
 
A large part of this paper is devoted to the answer of this question.
However, we also have studied what will happen for a homogeneous
extended BEC interacting with a weak running laser beam. Surprisingly,
non-Markovian effects similar to that in a PBG do occur even in this
non-periodic situation. This happens because in the presence of such a BEC
photons and excited atoms do form superpositions called
polaritons \cite{shlyapnikov91}.
The spectrum of these polaritons contains an avoided crossing
which has a similar effect on the SE rate as a PBG.

The paper is organized as follows. In Sec.~\ref{tmodel} we will 
present the theoretical
model on which our calculations are based. The general derivation of the SE
rates in laser fields will be done in Sec.~\ref{generalResult}. 
The results for the case of a BEC in a traveling wave laser beam or 
in a 1D optical lattice beam are discussed in Secs.~\ref{resultsRun} 
and \ref{resultsStand}
respectively, and are summarized in Sec.~\ref{conclusion}.
The details of the calculations are given in two Appendices. 

\section{The theoretical model}\label{tmodel}
We consider a BEC composed of two-level atoms which is coupled to the 
electromagnetic field. The interaction is described by using 
minimal coupling in
rotating-wave approximation under neglection of the term quadratic in the 
electromagnetic field. The interaction Hamiltonian is then given by
\begin{equation} H_{\mbox{{\scriptsize int}}} = 
      \int d^3 k d^3 k^\prime  \zeta_\sigma(\vec{k}) 
      a_\sigma(\vec{k}) \Psi_g(\vec{k}^\prime) \Psi_e^\dagger (\vec{k}+
      \vec{k}^\prime) +\mbox{H.c.} \; ,
\label{intham}
\end{equation}
with $\zeta_\sigma(\vec{k}) := \omega_{\mbox{{\scriptsize res}}} \vec{d}\cdot 
\vec{\varepsilon}_\sigma(\vec{k}) [\hbar/(2(2\pi)^3\varepsilon_0 \omega_k
]^{1/2}$ for an electromagnetic mode with frequency
$\omega_k=c|\vec{k}|$ and polarization vector
$\vec{\varepsilon}_\sigma(\vec{k})$. The vector $\vec{d}$ denotes
the atomic dipole moment and $\omega_{\mbox{{\scriptsize res}}}$
is the resonance frequency.
The Heisenberg equations of motion for the photon annihilation operators 
$a_\sigma(\vec{k})$ and the field operators $\Psi_e$
and $\Psi_g$ for excited and ground-state atoms can be derived easily and
are given by
\begin{eqnarray} 
  i \hbar \dot{\Psi}_e(\vec{k}) &=& \left \{ \frac{\hbar^2 \vec{k}^2}{2M}
  +\hbar \omega_{\mbox{{\scriptsize res}}} \right \} \Psi_e(\vec{k}) +
   \int d^3k^\prime \sum_\sigma
  \zeta_\sigma(\vec{k}^\prime ) a_\sigma(\vec{k}^\prime ) \Psi_g (
  \vec{k}-\vec{k}^\prime ) 
  \label{edgl} \\
  i \hbar \dot{\Psi}_g(\vec{k}) &=&  \frac{\hbar^2 \vec{k}^2}{2M}
  \Psi_g(\vec{k}) +
  \int d^3k^\prime \sum_\sigma
  \zeta^*_\sigma(\vec{k}^\prime ) a^\dagger_\sigma(\vec{k}^\prime ) 
  \Psi_e (\vec{k}+\vec{k}^\prime ) 
  \label{gdgl} \\
  i \hbar \dot{a}_\sigma(\vec{k}) &=&  \hbar \omega_k
  a_\sigma(\vec{k}) + \zeta_\sigma(\vec{k}) \int d^3k^\prime 
   \Psi_g^\dagger(\vec{k}^\prime ) \Psi_e (\vec{k}+\vec{k}^\prime ) 
\label{adgl}
\end{eqnarray} 
We have neglected the interatomic interaction terms.

To address the question of self-stabilization consider the following
situation: the atoms in the internal 
ground-state have formed a BEC which is described by a macroscopically
occupied coherent collective wavefunction 
$\Psi_g^{\mbox{{\scriptsize coh}}}$. They interact with a 
traveling wave or standing wave laser which is described by 
a coherent c-number field
$a_\sigma^{\mbox{{\scriptsize coh}}}(\vec{k})$.
Due to this interaction a part of the BEC is coherently excited.
We denote the wavefunction for coherently excited atoms by
$\Psi_e^{\mbox{{\scriptsize coh}}}$. Since both the ground-state BEC
and the laser beam are described by c-number fields it is easy to see
from Eq.~(\ref{edgl}) that $\Psi_e^{\mbox{{\scriptsize coh}}}$ must
be a c-number field, too. It is only through the spontaneous decay
of these coherently excited atoms that q-number deviations from 
c-number solutions to Eqs.~(\ref{edgl}) to (\ref{adgl}) can appear.
The corresponding SE rate determines the stability of the
macroscopic solution.

Let us start with the assumption that the BEC can indeed stabilize itself
against SE. In that case a
stationary macroscopic solution $(\Psi_g^{\mbox{{\scriptsize coh}}} ,
\Psi_e^{\mbox{{\scriptsize coh}}}, 
a_\sigma^{\mbox{{\scriptsize coh}}})$ of Eqs.~(\ref{edgl}) to (\ref{adgl})
should exist. The problem then can be
divided into two separate parts. We first search for 
the stationary  macroscopic coherent solution which includes all
interaction effects between atoms and photons beside SE.
Having found this solution we can perform a stability 
analysis to analyze the quantum fluctuations (SE) around it.
Spontaneous decay will make the coherent solution unstable and 
the corresponding quantum corrections will become important on a 
time scale comparable to the atomic lifetime (which is to be calculated).
For times shorter than this lifetime the deviations from the
coherent solution will be small (i.e., there are only few non-condensed
atoms and non-laser photons). 

Given a stationary macroscopic solution of the Heisenberg equations of motion
the stability analysis can be performed by applying Bogoliubov's method. 
This is done by writing the quantum field operators in the form
\begin{eqnarray} 
   \Psi_g(\vec{k})& =& \exp[-i \mu t] \{ \Psi_g^{\mbox{{\scriptsize coh}}}
   (\vec{k}) + \delta \Psi_g(\vec{k}) \} \label{fluctg} \\
   \Psi_e(\vec{k}) &=& \exp[-i (\mu+\omega_L) t] \{ 
   \Psi_e^{\mbox{{\scriptsize coh}}}(\vec{k}) + 
   \delta \Psi_e(\vec{k}) \} \label{flucte} \\
   a_\sigma(\vec{k}) &=& \exp[-i \omega_L t] \{ 
   a_\sigma^{\mbox{{\scriptsize coh}}}(\vec{k}) + \delta a_\sigma(\vec{k})\}
\label{flucta} \end{eqnarray}  
and retaining in Eqs.~(\ref{edgl}) to (\ref{adgl}) only terms linear in 
$\delta \Psi_i$ and $\delta a_\sigma$,
which describe the quantum fluctuations around the coherent solution.

The resulting linearized equations of motions are given by
\begin{eqnarray} 
  i\hbar \delta \dot{\Psi}_e(\vec{k}) &=& -\hbar \Delta_L 
  \delta \Psi_e(\vec{k})+
  \int d^3k^\prime \Psi_g^{\mbox{{\scriptsize coh}}}
  (\vec{k}\!-\!\vec{k}^\prime)   \delta a 
  (\vec{k}^\prime ) \zeta (\vec{k}^\prime ) 
  \label{fludgle} \\
  i\hbar \delta \dot{\Psi}_g(\vec{k}) &=& 
  \int d^3k^\prime \Psi_e^{\mbox{{\scriptsize coh}}}
  (\vec{k}+\vec{k}^\prime)   \delta a^\dagger  
  (\vec{k}^\prime ) \zeta (\vec{k}^\prime ) 
  \label{fludglg} \\
  i\hbar \delta \dot{a}(\vec{k}) &=& \hbar(c|\vec{k}| -\omega_L)
  \delta a (\vec{k}) + \zeta (\vec{k}) \int d^3k^\prime
  \left \{
  \Psi_e^{\mbox{{\scriptsize coh}}}(\vec{k}+\vec{k}^\prime) 
  \delta \Psi_g^\dagger (\vec{k}^\prime ) +
  \delta \Psi_e(\vec{k}+\vec{k}^\prime)  \Psi_g^{\mbox{{\scriptsize coh}}*} 
  (\vec{k}^\prime )
  \right \} 
 \label{fludgla} 
\end{eqnarray} 
Here  $\omega_L$ is the laser's frequency and 
$\Delta_L := \omega_L -\omega_{\mbox{{\scriptsize res}}}$ its detuning.
The linearized equations are valid as long as the photon-atom quantum 
fluctuations remain small enough, i.e.,
there are only few non-condensed atoms and non-laser photons.
This is certainly the the case for short times.
Furthermore several other approximations have been made.
First, it is not difficult to see that for a BEC of density $10^{14}$
cm$^{-3}$ the chemical potential $\hbar\mu$, the kinetic energy $\hbar^2
\vec{k}^2/(2M)$ of an atom, and the laser's Rabi frequency
$\Omega = a^{\mbox{{\scriptsize coh}}} \zeta_{\sigma_0}(\vec{k}_L)/\hbar$ 
are typically much smaller
than the interaction energy $\zeta_\sigma(\vec{k}) 
\Psi_g^{\mbox{{\scriptsize coh}}}(\vec{k})$
if $|\vec{k}|$ is of the order of $\omega_{\mbox{{\scriptsize res}}}/c$. We 
thus have neglected
all terms in which these quantities do appear.

\begin{figure}[t]
\epsfxsize=7cm
\hspace{1cm}
\epsffile{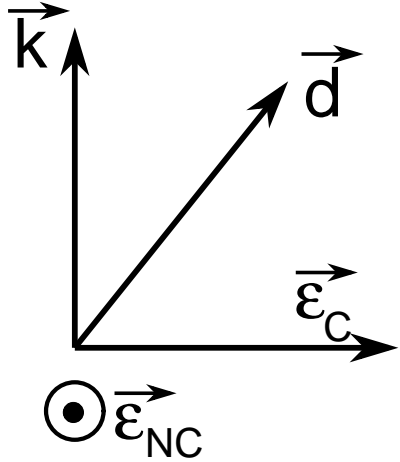}\vspace{-5mm}
\caption{\label{polvecs} The polarization vectors can be chosen
in a way that only one of them, 
$\vec{\varepsilon}_{\mbox{{\scriptsize C}}} (\vec{k})$, 
is not orthogonal to $\vec{d}$. In this case the three vectors
$\vec{d}$, $\vec{k}$, and 
$\vec{\varepsilon}_{\mbox{{\scriptsize C}}} (\vec{k})$ are in the same plane.
The second polarization vector 
$\vec{\varepsilon}_{\mbox{{\scriptsize NC}}} (\vec{k})$ 
is perpendicular to this plane.}
\end{figure}

In addition, we have introduced two specific polarization vectors
for the electromagnetic field.
A ``non-coupled'' polarization vector
$\vec{\varepsilon}_{\mbox{{\scriptsize NC}}} (\vec{k})$, which is
perpendicular to the photon momentum $\vec{k}$ and the atomic dipole 
moment $\vec{d}$, and a coupled polarization vector
$ \vec{\varepsilon}_{\mbox{{\scriptsize C}}} (\vec{k})$, which is
perpendicular to $\vec{k}$ and $ \vec{
\varepsilon}_{\mbox{{\scriptsize NC}}} (\vec{k})$ (see Fig.~\ref{polvecs}).
Since the interaction is proportional to the scalar product
of the polarization vector and $\vec{d}$ only electromagnetic
modes with polarization $ \vec{\varepsilon}_{\mbox{{\scriptsize C}}}(\vec{k})$
do interact with the atoms. We associate with these modes
the quantum fluctuation operator $\delta a(\vec{k}) := \delta a_{\sigma = C}
(\vec{k})$. It is easy to see that the scalar product 
$ \vec{\varepsilon}_{\mbox{{\scriptsize C}}} (\vec{k}) \cdot \vec{d}$, 
which appears in the definition of $\zeta_\sigma(\vec{k})$, is given by
$|\vec{d}| \sin \vartheta_{\vec{k}}$, where $\vartheta_{\vec{k}}$
is the angle between  $\vec{k}$ and $\vec{d}$. For notational convenience
we have defined $\zeta(\vec{k}) := \zeta_{\sigma = C}(\vec{k})$. 

It is possible to derive Eqs.~(\ref{fludgle}) to (\ref{fludgla})
from an effective Hamiltonian for the quantum fluctuations, 
\begin{equation} 
  H_{\mbox{{\scriptsize fluct}}} =   H_{\mbox{{\scriptsize pol}}}
  +   H_{\mbox{{\scriptsize spont}}} \; .
\label{hfluct} \end{equation} 
The first part,
\begin{eqnarray} 
   H_{\mbox{{\scriptsize pol}}} &=& \hbar \int d^3k \left \{ 
   -\Delta_L  \delta \Psi_e^\dagger \delta \Psi_e 
   + (c|\vec{k}| - \omega_L) \delta a^\dagger \delta a  \right \}
   \nonumber \\ & & 
   +\int d^3k d^3 k^\prime \zeta(\vec{k})\Psi_g^{\mbox{{\scriptsize coh}}}
    (\vec{k}-\vec{k}^\prime)
    \left \{ \delta a(\vec{k}) \delta \Psi_e^\dagger (\vec{k}^\prime )
    + \delta a^\dagger(\vec{k}) \delta \Psi_e(\vec{k}^\prime ) \right \}\; ,
\label{hpol} \end{eqnarray} 
conserves the number of photons plus excited atoms,
\begin{equation} 
   N_{\mbox{{\scriptsize pol}}} = \int d^3k \{ \delta \Psi_e^\dagger 
   \delta \Psi_e + \delta a^\dagger \delta a \}\; .
\label{npol} \end{equation} 
The first integral in $H_{\mbox{{\scriptsize pol}}}$ describes the energy
of free incoherent photons and atoms. 
The second integral represents the excitation of atoms 
from the ground-state BEC and the reabsorption of
incoherent photons by the BEC. 
Its eigenmodes $ |\vec{q},r \rangle = {\cal P}_{\vec{q},r}^\dagger | 0 \rangle$
are generally superpositions of photons and excited atoms,
i.e., polaritons \cite{shlyapnikov91}. They are characterized by a
continuous, momentum-like quantum number $\vec{q}$ and discrete quantum 
numbers $r$ (see below) and can generally be written as
\begin{eqnarray}  
  {\cal P}_{\vec{q},r}^\dagger &=& \int d^3k \{ {\cal E}_{\vec{q},r}(\vec{k})
   \delta \Psi_e^\dagger (\vec{k}) +
   {\cal A}_{\vec{q},r}(\vec{k}) \delta a^\dagger (\vec{k}) \}
  \label{linsup} \\
  \delta \Psi_e(\vec{k}) &=& \int d^3q \sum_r {\cal E}_{\vec{q},r}(\vec{k})
       {\cal P}_{\vec{q},r} \label{linsupe} \\
  \delta a(\vec{k}) &=& \int d^3q \sum_r {\cal A}_{\vec{q},r}(\vec{k})
       {\cal P}_{\vec{q},r} \label{linsupa}
\end{eqnarray} 
The form of the expansion coefficients ${\cal E}_{\vec{q},r}(\vec{k}) ,
{\cal A}_{\vec{q},r}(\vec{k})$
depends on the particular physical situation and is derived 
for a traveling and standing-wave laser in
Appendix \ref{polaritonEigenmodesRun} and
\ref{polaritonEigenmodesStand}, respectively.
The second part of the effective Hamiltonian is given by
\begin{eqnarray} 
   H_{\mbox{{\scriptsize spont}}} &=&     
   \int d^3k d^3k^\prime \zeta(\vec{k}^\prime) 
   \Psi_e^{\mbox{{\scriptsize coh}}}(\vec{k}+\vec{k}^\prime)
   \left \{
   \delta a^\dagger (\vec{k}^\prime ) \delta\Psi_g^\dagger (\vec{k}) + 
   \delta a (\vec{k}^\prime ) \delta\Psi_g (\vec{k}) \right \}
\end{eqnarray} 
It does not conserve $ N_{\mbox{{\scriptsize pol}}}$ and 
describes the spontaneous decay of coherently excited atoms.
If this term would vanish the macroscopic coherent state would be
stable against spontaneous decay.

\section{General derivation of SE rates}\label{generalResult}
The stability analysis essentially comprises to solve the time
evolution of the polariton modes for relatively short times during which 
the occupation of the macroscopic coherent solution does not change very much. 
This will allow us to derive the initial SE rate of coherently
excited atoms.
We assume that initially all atoms and photons are in the state determined by
the macroscopic coherent solution,
or in other words, the quantized polariton field (photon-atom quantum 
fluctuations)
is initially in the vacuum $|0 \rangle$. 
This state then evolves under the action of the fluctuation Hamiltonian
(\ref{hfluct}) into the time dependent state  $|\psi(t)\rangle$.

To describe this time evolution we rewrite the polariton Hamiltonian
(\ref{hpol}) in the convenient form
\begin{equation} 
  H_{\mbox{{\scriptsize pol}}} = \int d^3q \sum_r \hbar (\omega_{\vec{q},r}
  -\Delta_L) {\cal P}_{\vec{q},r}^\dagger  {\cal P}_{\vec{q},r}\; ,
\end{equation} 
where $\omega_{\vec{q},r}-\Delta_L$ are the eigenfrequencies of 
$H_{\mbox{{\scriptsize pol}}}$. Using Eqs.~(\ref{linsupe}) and
(\ref{linsupa}) one also can derive
\begin{equation} 
  H_{\mbox{{\scriptsize spont}}} = \int d^3k \int d^3q \sum_r \left \{
  \delta \Psi_g(\vec{k}) {\cal P}_{\vec{q},r} g_{\vec{q},r}(\vec{k}) +
  \delta \Psi_g^\dagger (\vec{k}) {\cal P}_{\vec{q},r}^\dagger g_{\vec{q},r}^*
  (\vec{k}) \right \} 
\end{equation} 
with
\begin{equation} 
  g_{\vec{q},r}(\vec{k}) := \int d^3k^\prime \zeta(\vec{k}^\prime )
  \Psi_e^{\mbox{{\scriptsize coh}}} (\vec{k}+\vec{k}^\prime ) 
  {\cal A}_{\vec{q},r}(\vec{k}^\prime )\; .
\label{gqr} \end{equation} 

To describe the evolution of the state $|\psi(t)\rangle$ we make the 
following ansatz, which corresponds to the one-photon approximation,
\begin{equation} 
  |\psi(t)\rangle \approx R(t) |0 \rangle +
  \int d^3k \int d^3q \sum_{r} 
  S_{\vec{q},r}(\vec{k},t) {\cal P}_{\vec{q},r}^\dagger 
  \delta\Psi_g^\dagger(\vec{k})  \;  |0 \rangle  \; .
\end{equation} 
The Schr\"odinger equation $i \hbar | \dot{\psi} \rangle = 
H_{\mbox{{\scriptsize fluct}}}  | \psi\rangle $ then can be solved
by using the Laplace transform $\bar{R}(s) =
\int_0^\infty \exp[-ts] R(t) dt$ and similarly for
$S_{\vec{q},s}(\vec{k},t)$. The resulting equations,
\begin{eqnarray} 
  i\hbar (s \bar{R}(s)- R(0)) &=& \int d^3k \int d^3q \sum_r 
  \bar{S}_{\vec{q},r}(\vec{k},s) g_{\vec{q},r}(\vec{k}) \\
  i \hbar s  \bar{S}_{\vec{q},r}(\vec{k},s) &=& \hbar (\omega_{\vec{q},r}
  -\Delta_L)  \bar{S}_{\vec{q},r}(\vec{k},s) + \bar{R}(s) 
  g_{\vec{q},r}^*(\vec{k})  \; ,
\end{eqnarray} 
have the solution
\begin{equation} 
  \bar{R}(s) = \frac{ R(0)}{ s - I(s) }\; .
\label{lapltransf} \end{equation} 
The dependence on the particular physical situation is completely
determined by the integral 
\begin{equation} 
  I = \frac{1}{i\hbar^2}  \int d^3q \sum_r
  \frac{ \int d^3k |g_{\vec{q},r}(\vec{k})|^2 }{z_s-\omega_{\vec{q},r} } \; .
\label{integral} \end{equation}
For notational convenience we have defined the complex variable
\begin{equation} 
  z_s := i s + \Delta_L \; .
\end{equation} 

The most important aspect of $I$ is its
complex analytical structure. This is because the inverse Laplace transform
is defined by
\begin{equation} 
  R(t) = \frac{1}{2\pi i} \int_{\epsilon -i \infty}^{\epsilon +i \infty}
  e^{ts} \bar{R}(s) ds \; ,
\label{inverseLapl} \end{equation} 
where $\epsilon$ is chosen so that the path of integration lies to the right
of any branch cuts and poles of $\bar{R}(s)$.
From Eq.~(\ref{lapltransf}) it becomes clear that the branch cuts of
$\bar{R}(s)$ are those of  $I$ and that
the poles of $\bar{R}(s)$ essentially depend on the form of
$I$. Assuming that all poles of $\bar{R}(s)$
are simple poles we then find for $R(t)$
\begin{equation} 
  R(t) = \sum_{s_i} e^{t s_i} \mbox{Res}(\bar{R}(s), s_i) + \sum_{{\cal B}_j}
          \frac{1}{2\pi i} \int_{{\cal B}_j}e^{ts} \bar{R}(s) ds  \; ,
\label{atsol} \end{equation} 
where $s_i$ denote the poles of $\bar{R}(s)$ and ${\cal B}_j$ the
branch cuts. Each pole corresponds to a fraction of the coherently excited
atoms which decays (or increases exponentially) with a SE rate of
\begin{equation} 
 \gamma_i = -2\mbox{Re}(s_i). \label{poles}
\end{equation}
The integration contours around the branch cuts corresponds
to a fraction of coherently excited atoms with a non-exponential time 
evolution.

With Eq.~(\ref{poles}) we have found a general expression for the SE
rates that can appear in the presence of a BEC. We now want
to study the different physical situations of a BEC in a traveling or 
standing wave laser and to derive the corresponding values of $\gamma_i$.
To do so we have to find closed expressions for the polariton
eigenfrequencies $\omega_{\vec{q},r}$ and the functions
$g_{\vec{q},r}(\vec{k})$ in order to derive $I(s)$. These quantities
in turn require the knowledge of both the polariton eigenmodes and the
fields  $(\Psi_g^{\mbox{{\scriptsize coh}}} ,
\Psi_e^{\mbox{{\scriptsize coh}}}, a_\sigma^{\mbox{{\scriptsize coh}}})$ 
comprising the macroscopic coherent 
solution. Since the calculations leading to a closed expression for
$I$ are quite involved we present them in the appendices.  
In the next two sections we analyse the results and give physical
interpretations of the effects involved. 

\section{Spontaneous emission rates for a BEC in a traveling wave 
laser} \label{resultsRun}
\begin{figure}[t]
\epsfxsize=10cm
\hspace{1cm}
\epsffile{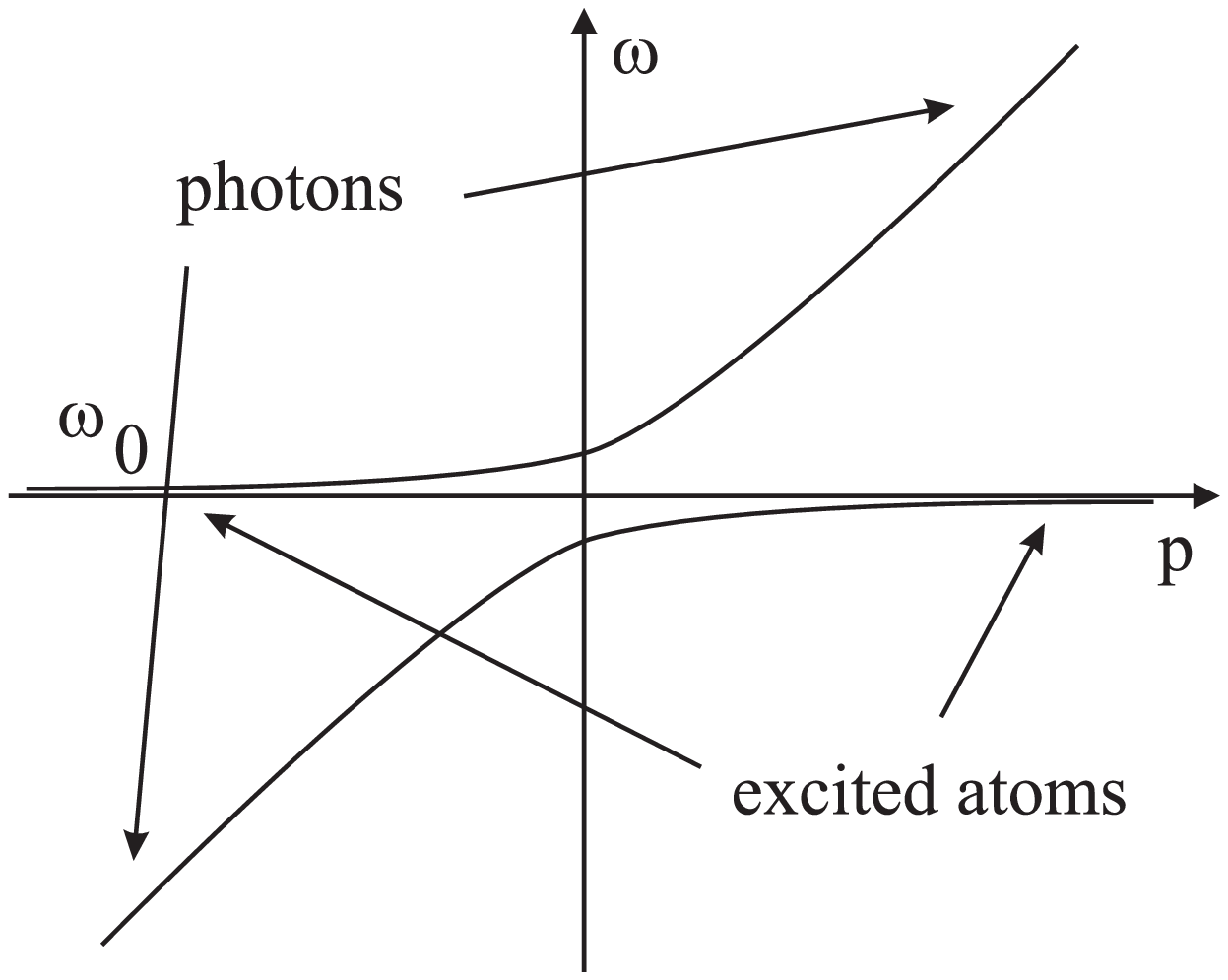}\vspace{-5mm}
\caption{\label{avoidedCrossing} A homogeneous BEC induces an avoided 
crossing in the polariton spectrum.
Far away from the avoided crossing the polaritons describe excited
atoms or photons. Thus, if one focuses on the photons, the avoided crossing
provides an effective band gap.}
\end{figure}

In the case of a BEC interacting with a traveling wave laser,
the polariton dispersion relation derived 
in Appendix \ref{polaritonEigenmodesRun}
does contain an avoided crossing around the resonance
frequency of the atoms (see Fig.~\ref{avoidedCrossing}). 
Since nearly resonant photons provide the  dominant 
contribution to SE, it is physically evident that this avoided crossing in
the dispersion spectrum will produce an effect on the SE rate which is similar
to what a PBG can do. 

This effect can be studied by analysing the closed expression for
the renormalized value of $I(s)$ which we have derived using
the generalized Wigner-Weisskopf approximation presented in
Appendix \ref{derivationOfIRun}. We find 
\begin{equation}
   I^{\mbox{{\scriptsize Ren}}}  =  
   \left ( 1 - \frac{4\nu_g}{5 z_s }
   \right )  I_0^{\mbox{{\scriptsize Ren}}} +
   \frac{ N_e \gamma_{\mbox{{\scriptsize vac}}}}{5\pi i} 
   \left \{- \frac{47 \nu_g}{15 z_s} 
   + {8\over 3}  + \frac{z_s}{\nu_g}
   + \left ( 1+{4\nu_g\over z_s}\right )
   (1-{ z_s\over \nu_g})^{3/2} \mbox{arcoth}
   (\sqrt{1-{z_s\over \nu_g}}) \right \}
\label{iren}
\end{equation}
In this expression 
$\gamma_{\mbox{{\scriptsize vac}}} := 
\vec{d}^2 \omega_{\mbox{{\scriptsize res}}}^3/(3\pi\hbar \varepsilon_0 c^3)$
denotes the SE rate in free space and $N_e := V \rho_e$
the number of coherently excited atoms ($V$ is the quantization volume).
The frequency
\begin{equation} 
  \nu_g := \frac{ |\vec{d}|^2 \rho_g }{2\hbar \varepsilon_0}\; ,
\label{nug} \end{equation} 
with $\rho_g$ being the density of atoms in the ground-state,
determines the strength of the interaction between photons and
excited atoms mediated by the BEC. Typically we have
$E_{\mbox{{\scriptsize interaction}}} = 
\hbar \sqrt{\nu_g \omega_{\mbox{{\scriptsize res}}}}$
(see appendix \ref{polaritonEigenmodesRun}).

As already mentioned in Sec.~\ref{generalResult} the time evolution
of the macroscopic coherent solution depends essentially on the analytical
structure of $\bar{R}(s)$ of Eq.~(\ref{lapltransf}). 
In general, $\bar{R}(s)$ has several poles and
a branch cut originating from the term including the arcoth in
$I^{\mbox{{\scriptsize Ren}}}$. This cut lies between $z_s= 0$ and
$z_s=\nu_g$. 
Another important property of
$I^{\mbox{{\scriptsize Ren}}}$ is that it depends on $z_s$ only through
the ratio $z_s/\nu_g$ so that
the magnitude of the SE modification depends on this ratio, too.
In addition, a numerical evaluation of Eq.~(\ref{iren}) shows that
it is a slowly varying function of the order 
$\gamma_{\mbox{{\scriptsize vac}}} N_e$ unless $z_s/\nu_g$ is close to
zero. This has the following consequences.

In free space $z_s$ can be taken to be close to the pole
of $\bar{R}(s)$, i.e., of the order of
$\gamma_{\mbox{{\scriptsize vac}}} N_e$. 
Hence, the magnitude of non-Markovian effects is essentially determined
by the ratio
\begin{equation} 
  \frac{z_s}{\nu_g} \approx  
  {N_e \gamma_{\mbox{{\scriptsize vac}}} \over \nu_g} = 
  \frac{16\pi^2}{3} \frac{ N_e}{\rho_g \lambda^3} \; .
\end{equation} 
This ratio is proportional to the total number of excited atoms divided 
by the number of condensed ground-state atoms per optical wavelength 
$\lambda$. The request that this ratio should be small 
has important consequences when applied to BECs
of a density in the order of $10^{14}$ cm$^{-3}$. In this case the
BEC-induced effects can only be relevant if there are very few
coherently excited atoms (in the order of one). However, for higher densities
significant effects can occur also for a higher number of
excited atoms. 

The request that $N_e$ is of order one
also implies $|\Delta_L| \gg \Omega_L$ since otherwise
the Rabi frequency would be large enough to excite many atoms.
As discussed in Appendix \ref{classicalSolutionRun}, 
the macroscopic coherent
solution implies in this case for $\Delta_L>0$ the additional
constraint $\Delta_L >4\nu_g$.
In Fig.~\ref{resultsFigRun} the real part of the two dominating 
poles $s_1, s_2$ of $I(s)$ is shown
as a function of $\Delta_L$ for the case $N_e=1$ and  $\nu_g = 2.5 
\gamma_{\mbox{{\scriptsize vac}}}$
(corresponding to an atom density $\rho_g$ of $5\times 10^{14}$ cm$^{-3}$).
For $\Delta_L>0$ a third pole appears with a very small negative decay rate
($<10^{-3} \gamma_{\mbox{{\scriptsize vac}}}$).
The occurrence of negative decay rates is consistent within the range of
validity of the linearized equations for the quantum fluctuations
and may indicate the formation of an atom-photon bound state
\cite{john90}.
Obviously the change in the SE rate can be quite large for small $|\Delta_L|$.
According to Eq.~(\ref{atsol}) the fraction of atoms belonging to the
poles can be easily calculated by determining the residue at the poles.
It turns out that the pole whose real part asymptotically approaches
$\gamma_{\mbox{{\scriptsize vac}}}$ always dominates and that the fraction of 
atoms belonging to other
poles is significant only for small $|\Delta_L|$. The same holds for the
fraction corresponding to the branch cut.

If $|\Delta_L| \gg \gamma_{\mbox{{\scriptsize vac}}}$ holds the dominant pole 
$s_1$ can be calculated by perturbation theory. Its real part (the decay rate)
is then given by
\begin{equation} 
  {1\over 2} \gamma (\Delta_L)  =  N_e  {\gamma_{\mbox{{\scriptsize vac}}}
  \over 2}   
  \left ( 1- {4 \nu_g \over 5 \Delta_L} \right )
  +O(\Delta_L^{-2})   \; .
\label{homogExpan} \end{equation} 
We see that the SE rate is altered by a factor of $1- 4\nu_g/(5\Delta_L)$. 
It depends on the sign of $\Delta_L$ whether SE is increased or decreased.
We remark that the reason why SE depends on the detuning is that the
coherently excited atoms are driven by the laser field and thus oscillate
at the laser frequency $\omega_L$ instead of the resonance frequency
$ \omega_{\mbox{{\scriptsize res}}}$ (see Eq.~(\ref{flucte})).

We shortly summarize the results that we have found for a BEC interacting
with a running laser beam. For evolution times smaller than the atomic
lifetime and for a weak laser beam which only excites a number 
$N_e = O(1)$ of excited atoms (for BEC densities of $10^{14}$ cm$^{-3}$),
the spontaneous emission rate is significantly modified by
non-Markovian effects. These effects result from an avoided
crossing in the polariton spectrum caused by 
the extended homogeneous BEC in the internal ground state. 

In the next section we will examine the corresponding results for a spatially
periodic (lattice) BEC in a standing wave laser beam.

\section{Spontaneous emission of a BEC in a standing wave laser} 
\label{resultsStand}
A BEC in an optical lattice formed by a standing wave laser beam
becomes spatially periodic and thus provides
a kind of periodic dielectric. The polariton spectrum will therefore
contain band gaps. We have derived the corresponding
dispersion relation in Appendix \ref{polaritonEigenmodesStand}. As is
well known from PBGs a band gap around the resonance frequency will
lead to non-Markovian effects in the spontaneous emission of an atom.
The examination of these effects will answer the question whether a
periodic BEC in an optical lattice can stabilize itself against SE.

To determine the SE rate of a BEC in a 1D optical lattice we have again 
to evaluate the integral $I$ of Eq.~(\ref{integral}). This task is quite
involved and is presented in appendix \ref{derivationOfIStand}. 
Our final analytical form of the renormalized integral 
$I^{\mbox{{\scriptsize Ren}}}$ is given by the somewhat lengthy expression
\begin{eqnarray} 
    I^{\mbox{{\scriptsize Ren}}} &=& 
    \frac{\bar{N}_e \gamma_{\mbox{{\scriptsize vac}}}}{2\pi i} 
    \left \{ 2 + \ln \left (
	\frac{\Lambda}{\omega_{\mbox{{\scriptsize res}}}} \right ) 
    \right \} +
  \label{analyticalResult} \\ & &
    \frac{ i\gamma_{\mbox{{\scriptsize vac}}} \bar{N}_e}{\pi} 
    \left \{  
       \int_1^\infty v dv
       \ln \left [ \frac{(f_0(z_s)-v)(\sqrt{v^2+1}-f_0(z_s))+f_1(z_s)  }{
                   (f_0(z_s)-v)(\sqrt{v^2-1}-f_0(z_s))+f_1(z_s)        }
           \right ]  + 
    \right .
  \nonumber \\ & & \hspace{1.2cm} 
    \left .
        \int_0^1 v dv
         \ln \left [ \frac{(f_0(z_s)-v)(\sqrt{v^2+1}-f_0(z_s))+f_1(z_s)  }{
                     (f_0(z_s)-v)(1-v-f_0(z_s))+f_1(z_s)                 }
             \right ] 
    \right \} +
  \nonumber \\ & &
    \frac{i\gamma_{\mbox{{\scriptsize vac}}} 
      \tilde{N}_e}{\pi} { \tilde{\nu}_g\over z_s}
      \int _1^\infty \frac{\sqrt{v} dv}{f_0(z_s)-v} 
    \Bigg \{ 
            (v^2+1)^{1/4} -(v^2-1)^{1/4} - 
  \nonumber \\ & & \hspace{2cm}  
            h(v)\mbox{arctanh} \left[\frac{(v^2+1)^{1/4}}{h(v)} \right ] + 
            h(v)\mbox{arctanh} \left[\frac{(v^2-1)^{1/4}}{h(v)} \right ]
    \Bigg \}+
  \nonumber \\ & & 
    \frac{i\gamma_{\mbox{{\scriptsize vac}}} \tilde{N}_e}{\pi}  
    {\tilde{\nu}_g\over z_s} \int_0^1 \frac{\sqrt{v} dv}{f_0(z_s)-v} 
    \Bigg \{ 
           (v^2+1)^{1/4} -(1-v)^{1/2} -      
  \nonumber \\ & & \hspace{2cm} 
           h(v)\mbox{arctanh}\left[\frac{(v^2+1)^{1/4}}{h(v)} \right ] + 
           h(v)\mbox{arctanh}\left[\frac{\sqrt{1-v}}{h(v)}    \right ]
    \Bigg \}   \; .
 \nonumber 
\end{eqnarray} 
In this result we have introduced a couple of new notations. 
For notational convenience we have defined
\begin{eqnarray} 
    f_0(z_s) &:=& \frac{z_s +\omega_{\mbox{{\scriptsize res}}}}{2 c k_L} 
    -\frac{\omega_{\mbox{{\scriptsize res}}}}{
    2 c k_L} {\bar{\nu}_g\over z_s}  
  \label{f0} \\
    f_1(z_s) &:=& \frac{ \tilde{\nu}_g}{z_s} 
    \frac{\omega_{\mbox{{\scriptsize res}}} }{2 c k_L}  
\label{f1} \end{eqnarray} 
as well as the abbreviation
$h(v) := \sqrt{f_0(z_s) -f_1(z_s)/(f_0(z_s)-v)}$. We also introduced a cut-off
$\Lambda \approx m_e c^2/\hbar$ to regularize the integral ($m_e$ is the
electron's mass).
Two important physical quantities are given by
\begin{eqnarray} 
  \bar{N}_e &:=& {V\over(2\pi)^3} \sum_m 
   (\Psi_{e,2m+1}^{\mbox{{\scriptsize coh}}})^2 \label{barNe} \\
  \tilde{N}_e&:=& {V\over(2\pi)^3} \sum_m 
  \Psi_{e,2m-1}^{\mbox{{\scriptsize coh}}}
  \Psi_{e,2m+1}^{\mbox{{\scriptsize coh}}}
\label{tildeNe} \end{eqnarray}  
where $V$ denotes the quantization volume and the sum runs over the
(real) momentum components $\Psi_{e,m}^{\mbox{{\scriptsize coh}}}$ 
of coherently excited atoms.
$\bar{N}_e$ is simply the total number of excited atoms in the
macroscopic coherent field, and $\tilde{N}_e$ describes how these
atoms are distributed in momentum space and is always smaller than
$\bar{N}_e$. It is a measure for the degree of periodicity of the
density of excited atoms, very roughly we have 
$V \rho_e(z) \approx \bar{N}_e + \tilde{N}_e \cos(2 z k_L)$.

The influence of the
BEC in a standing wave laser on the SE rate is determined by the
frequencies
\begin{eqnarray} 
   \bar{\nu}_g &:=& {\zeta^2(\vec{k}_L)\over \hbar^2 
	\omega_{\mbox{{\scriptsize res}}}} \bar{\rho}_g(2\pi)^3 
         \approx \frac{\bar{\rho}_g \vec{d}^2}{2\hbar \varepsilon_0}
   \label{barnug} \\
   \tilde{\nu}_g&:=&{\zeta^2(\vec{k}_L)\over \hbar^2
	\omega_{\mbox{{\scriptsize res}}}}  \tilde{\rho}_g(2\pi)^3
         \approx \frac{\tilde{\rho}_g \vec{d}^2}{2\hbar \varepsilon_0} \; ,
\label{tildenug} \end{eqnarray} 
In the polariton dispersion relation $\bar{\nu}_g$ produces a
contribution similar to that of $\nu_g$ in the case of a BEC in a traveling
wave laser beam (avoided crossing). $\tilde{\nu}_g$ produces
a real PBG close to the resonance frequency
due to the spatial periodicity of a BEC in a 1D optical lattice.
The two frequencies define the strength of the
interaction mediated by the mean density $\bar{\rho}_g$ and 
the periodic part $\tilde{\rho}_g$ of the ground-state BEC,
\begin{eqnarray} 
  \bar{\rho}_g &:=& {1\over(2\pi)^3} \sum_m 
        (\Psi_{g,2m}^{\mbox{{\scriptsize coh}}})^2 
  \label{barrhog} \\
  \tilde{\rho}_g&:=& {1\over(2\pi)^3} \sum_m 
   \Psi_{g,2m}^{\mbox{{\scriptsize coh}}}
   \Psi_{g,2m+2}^{\mbox{{\scriptsize coh}}} \; .
\label{tilderhog} \end{eqnarray}  
These densities
play a similar role to what $\tilde{N}_e/V$ and $\bar{N}_e/V$ do for 
coherently excited atoms. For a mean density of
$\bar{\rho}_g \approx 10^{14}$ cm$^{-3}$ we find $\bar{\nu}_g \approx
4 \times 10^{6}$ Hz. The magnitude of $\tilde{\nu}_g$ can vary between
$\bar{\nu}_g$ for a very strong optical potential and 0 if the laser beam
is switched off.

Although Eq.~(\ref{analyticalResult}) has a complicated structure
it allows to analyze the main features of $I^{\mbox{{\scriptsize Ren}}}$
and hence of the time evolution of the macroscopic coherent solution
in the presence of (small) quantum fluctuations. It is even possible
to estimate the influence of the band gap with some simple arguments.
\subsection{General structure of the result}
A very important feature of the integral (\ref{analyticalResult})
is that all parts of $I^{\mbox{{\scriptsize Ren}}}$
are proportional to $\bar{N}_e \gamma_{\mbox{{\scriptsize vac}}}$ or 
$\tilde{N}_e \gamma_{\mbox{{\scriptsize vac}}}$.
In addition, it becomes obvious that $I^{\mbox{{\scriptsize Ren}}}$
depends on the ground-state BEC and on the complex variable
$z_s$ essentially through the ratios
$z_s/\bar{\nu}_g$ and $z_s/\tilde{\nu}_g$. The only exception to this
is the first term containing $z_s$ in Eq.~(\ref{f0}), but this term
is negligible compared to $\omega{\mbox{{\scriptsize res}}}$ and does
only serve to keep book on which side of the branch cut $z_s$ is placed
(see remarks below Eq.~(\ref{freeContrib})).

These facts can be exploited to estimate under which circumstances the
influence of the BEC on the SE rate is significant. Since for a BEC in a
traveling laser beam the contribution of the poles of $\bar{R}(s)$
usually dominates (see Sec.~\ref{resultsRun}), we will focus on this part.
The denominator of  $\bar{R}(s)$ is of the form
$s- I^{\mbox{{\scriptsize Ren}}}(s) $, see Eq.(\ref{lapltransf}). 
Since a numerical analysis of Eq.~(\ref{analyticalResult}) shows that  
$ I^{\mbox{{\scriptsize Ren}}}(s)$ is of the order of its pre-factors
$\bar{N}_e \gamma_{\mbox{{\scriptsize vac}}}$ or $\tilde{N}_e 
\gamma_{\mbox{{\scriptsize vac}}}$
unless $z_s/\bar{\nu_g}$ or  $z_s/\tilde{\nu_g}$ are small,
a pole $s_i$ must be of
the order of these pre-factors (if the detuning $\Delta_L$ is not very large).
In analogy to the case studied in Sec.~\ref{resultsRun}
one can again infer that the magnitude of the BEC-induced effects essentially
depends on ratios of the form 
$\bar{N}_e \gamma_{\mbox{{\scriptsize vac}}}/\bar{\nu}_g$, for instance. 
As in Sec.~\ref{resultsRun} this allows the conclusion that for a BEC
with a density in the order of $10^{14}$ cm$^{-3}$ non-Markovian effects
can only be relevant if there are very few
coherently excited atoms (in the order of one).

In the case of a BEC in an optical lattice this restriction has additional
implications: the Rabi frequency $\Omega$ of the coherent standing wave
laser has to be very small since otherwise too many atoms would be excited.
The number of excited atoms is approximately given by
$\bar{N}_e \approx (\Omega/\Delta_L)^2 \bar{N}_g$, where $\bar{N}_g$
denotes the total number of condensed ground-state atoms. Since
$\bar{N}_g$ is very large the ratio $\Omega/\Delta_L$ must be very small
in order to achieve $\bar{N}_e \approx 1$. This, in turn, means that
the optical potential ($\propto \Omega^2/\Delta_L$)
provided by the standing laser beam is very weak
and thus $\tilde{\nu}_g $ is much smaller than  $\bar{\nu}_g $.
A small value of $\tilde{\nu}_g $ simply means that the polaritonic
band gap that is formed in the presence of a BEC will be small
and therefore will not have a significant effect on the SE rate. 

\subsection{Interference channel for spontaneous emission in PBG}
Another observation deals with the dependence of
$I^{\mbox{{\scriptsize Ren}}}$ on the wavefunction 
$\Psi_e^{\mbox{{\scriptsize coh}}}$ of
coherently excited atoms. It is known that in free space
the shape of the spatial wavefunction of an excited atom does only have
a tiny influence on its SE rate \cite{rzazewski92}. These
small corrections are mainly due to the atomic kinetic energy 
which we have neglected in the Hamiltonian (\ref{hfluct}) for the quantum
fluctuations. 
In this sense, one would expect that the SE rate in 
Eq.~(\ref{analyticalResult}) does also not depend on the shape of the
wavefunction for coherently excited atoms and is
proportional to total number $\bar{N}_e$ of excited atoms 
in which this shape does not enter. 
However, Eq.~(\ref{analyticalResult}) does also include the terms 
proportional to the quantity $\tilde{N}_e$
which clearly depend
on the shape (for instance, $\tilde{N}_e$ vanishes for a 
spatially homogeneous wavefunction $\Psi_e^{\mbox{{\scriptsize coh}}}$). 
Principally, this
contribution can be as large as that depending on $\bar{N}_e$.

This new dependence on the shape of $\Psi_e^{\mbox{{\scriptsize coh}}}$
is an additional effect of the 
BEC on the SE rate and no consequence of the polaritonic dispersion
relation. To understand its origin it is useful to look at
Eq.~(\ref{betasum}) where $\tilde{N}_e$ appears first. 
This contribution obviously does vanish if  ${\cal A}_0(\vec{q},r)$
or  ${\cal A}_{-1}(\vec{q},r)$, which are momentum-components of the photon
modes belonging to momentum $\vec{q}$ and $\vec{q}-2\vec{k}_L$, is zero. 
This is the case for photons interacting with a homogeneous BEC, for instance.
The new effect therefore can be considered as arising from the interference
between different momentum-components of the photon
modes and the wavefunction of coherently excited atoms.

We want to emphasize that this effect is not tied to the presence of a BEC.
The only conditions for its existence are the periodicity of both the
wavefunction $\Psi_e^{\mbox{{\scriptsize coh}}}$
for excited atoms and the eigenmodes for the photons.
Since in an ordinary PBG material the photon eigenmodes are periodic, 
this new contribution to the SE rate can be present in ordinary PBG
materials, too. In an ordinary periodic dielectric
the new interference channel for SE even 
could produce large contributions since the
periodicity of the dielectric is produced by, e.g., mechanical forces 
but not by the light that is used to excite the atoms. 
Only in the case of a BEC do the standing laser beams play a double
role, excitation of atoms and production of a periodicity in the BEC,
which results in a suppressed influence of both the polaritonic band gap and
the interference channel on the SE rate.

\section{Conclusion}\label{conclusion}
We conclude this paper by summarizing the results that we have found.
We have examined the self-stabilization of a BEC against SE by
performing a stability analysis of a macroscopically occupied state
for photons and two-level atoms, which
describes a BEC that is coherently coupled to a laser beam.
The presence of the ground-state BEC thereby leads to the
formation of polaritons and introduces non-Markovian effects in the
spontaneous decay of excited atoms.

\begin{figure}[t]
\epsfxsize=10cm
\hspace{1cm}
\epsffile{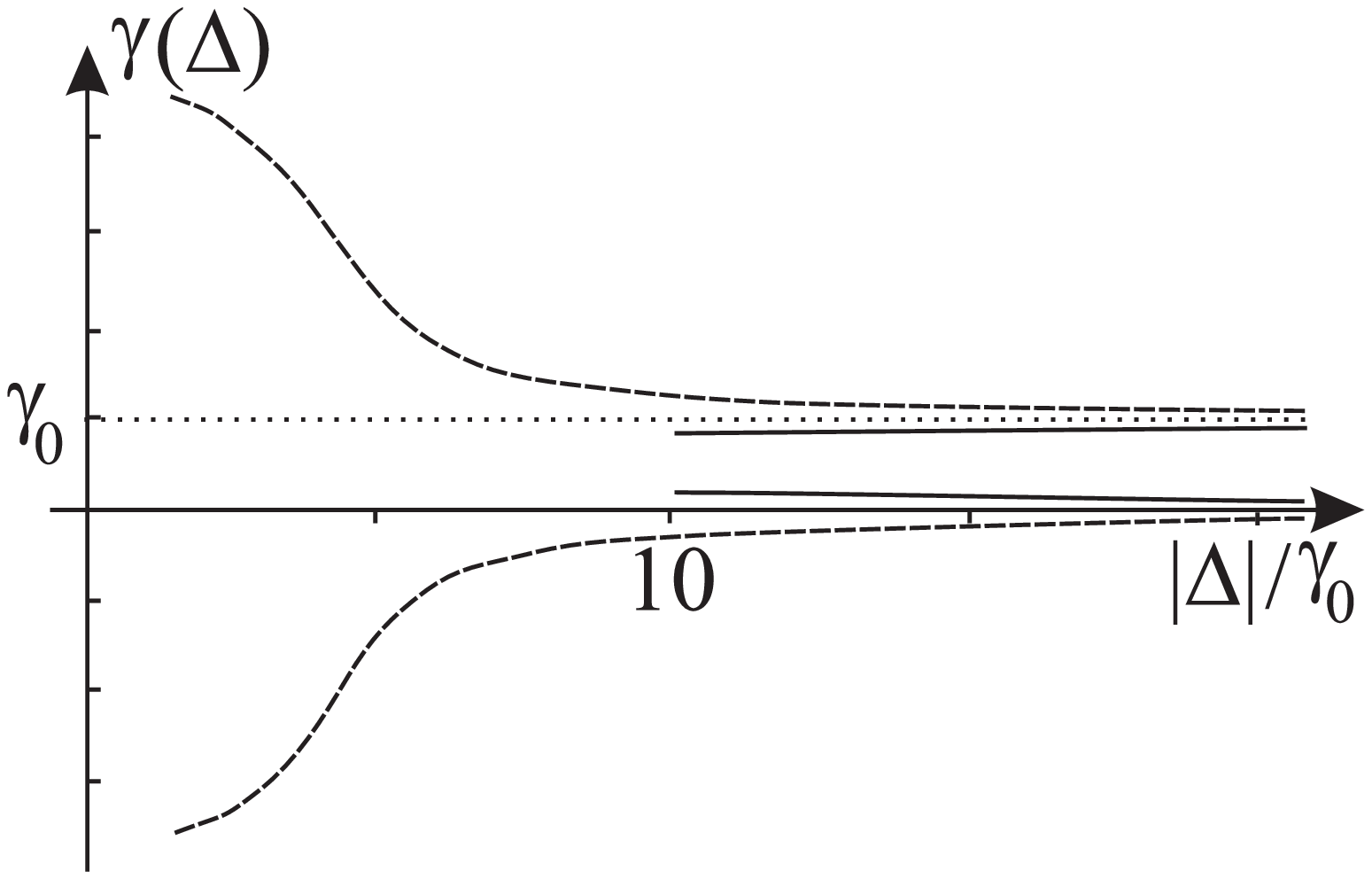}\vspace{-5mm}
\caption{\label{resultsFigRun} Spontaneous emission rate of a partially 
excited BEC in a running laser wave for detuning $\Delta_L>4 \nu_g$ 
(solid lines) and $\Delta_L<0$ (dashed lines). The atoms break up
into different fractions with different decay rates. The dominating fraction
is the one whose decay rate asymptotically approaches 
$\gamma_{\mbox{{\scriptsize vac}}}$.}
\end{figure}

In the case of a BEC in a traveling-wave laser, the polariton spectrum
displays an avoided crossing around the resonance frequency which causes
similar changes in the SE rate as a PBG in periodic dielectrics.
 Its magnitude depends
on the ratio $ N_e/(\rho_g \lambda_L^3)$
between the total number of excited atoms
$N_e$ and the number of BEC-atoms inside a cube
of the size of an optical wavelength $\lambda_L$. If this ratio is 
much larger than 1 the SE rate will essentially remain unchanged. 
Otherwise the change can be significant as the numerical examples shown in 
Fig.~\ref{resultsFigRun} demonstrate.
The change of the SE rate displayed in Fig.~\ref{resultsFigRun}  
depends on the detuning
of the laser because the coherently excited atoms are driven at the
laser's frequency $\omega_L$.

For a BEC in a 1D optical lattice two new effects do appear. Being
a kind of periodic dielectric the BEC then produces a real polaritonic 
band gap. The size of this band gap is determined by 
$\tilde{\nu}_g$. As in the
case of a traveling wave laser, SE is only significantly altered if
there are very few excited atoms. This in turn does imply that the
optical lattice must be very weak and therefore produces only a
small band gap which has only a very small influence on the SE rate.
The second new effect in a periodic BEC is the appearance of a new
channel for SE which arises from the interference between 
different momentum components of the excited-state wavefunction and the photon
modes. Though its effect in a BEC is as small as that of the band gap
it should also be present in the case of a PBG in an ordinary periodic
dielectric where it can be large.

It should be pointed out clearly where exactly the difference between
an ordinary periodic dielectric and a BEC in an optical lattice
comes into play. In an ordinary dielectric medium the periodicity
is produced by whatever forces determine the stability of the medium.
The excitation of an atom inside such a medium is done by a light
beam, i.e., a completely different physical system.
In the case of a BEC in an optical lattice, however, the excitation
of the atoms and the potential that produces the periodicity of the
BEC both are provided by the same device: the laser beams of the
optical lattice. These lattice beams have to achieve two competing
goals: to provide a strong periodic potential (to produce  a large
band gap) and to cause only a weak excitation (to have few excited atoms).
As the achievement of both goals is impossible the periodicity of
the BEC will only have a tiny influence on the SE rate and it will
essentially cause the same effect as a homogeneous BEC in a running
laser wave. 

This argument also provides the answer to the question whether 
self-stabilization of a BEC against SE is possible. Since the SE rate
is only significantly changed if there are very few excited atoms,
and since a large PBG does only form for strong laser beams,  
a self-stabilization is not possible for BECs with a density in the order
of $10^{14}$ cm$^{-3}$. 

We finally remark that our results are not applicable
to BECs confined in a micrometer-sized trap, a case discussed 
in the literature \cite{javanainen93,cooper95}. 
Our work is concerned with BECs which are extended enough to allow 
the formation of polaritons. The necessary extension of the BEC can be
estimated by considering the typical interaction energy 
for the formation of polaritons which is given by 
$\hbar \sqrt{\nu_g \omega_{\mbox{{\scriptsize res}}}}$ 
(see appendix \ref{polaritonEigenmodesRun}). For a BEC with a density
of $10^{14}$ cm$^{-3}$ this energy is in the order of $\hbar \times
10^{11}$ Hz. For the formation of polaritons a photon must therefore
be inside the BEC longer than $10^{-11}$ seconds. Since it travels at
the speed of light the BEC must therefore be larger than about 3 mm.


{\bf Acknowledgment}: 
This work has been supported by the Australian Research Council
and the Optik Zentrum Konstanz.

\begin{appendix}
\section{BEC in a running laser wave} \label{runwavAppendix}
\subsection{Derivation of the macroscopic coherent 
solution} \label{classicalSolutionRun}
We are interested in finding a particular solution
$(\Psi_g^{\mbox{{\scriptsize coh}}} , 
 \Psi_e^{\mbox{{\scriptsize coh}}},
 a_\sigma^{\mbox{{\scriptsize coh}}}(\vec{k}))$
of macroscopically occupied fields
to Eqs.~(\ref{edgl})-(\ref{adgl})
which describes a BEC coherently coupled to a 
running laser wave. We thus make the ansatz
\begin{eqnarray}
  \Psi_g^{\mbox{{\scriptsize coh}}} (\vec{k}) &=& 
       (2\pi)^{3/2} \sqrt{\rho_g} \delta(\vec{k})
                       \exp[-i\mu t] \label{cohg}\\
   a_\sigma^{\mbox{{\scriptsize coh}}}(\vec{k}) &=&
   \exp [-i \omega_L t] \delta(\vec{k}-\vec{k}_L) \delta_{\sigma,\sigma_L}
   \Omega_L [2(2\pi)^3 \hbar \varepsilon_0 \omega_{k_L}]^{1/2}/(|\vec{d}| 
   \omega_{\mbox{{\scriptsize res}}})  \label{coha}\\
   \Psi_e^{\mbox{{\scriptsize coh}}} (\vec{k}) &=& 
   (2\pi)^{3/2} \sqrt{\rho_e} \delta(\vec{k}-\vec{k}_L)
   \exp [-i (\mu+\omega_L) t] \label{cohe} 
\end{eqnarray} 
which corresponds to a homogeneous ground-state BEC of density $\rho_g$,
a laser beam with frequency $\omega_L$, Rabi frequency
$\Omega_L>0$, polarization $\sigma_L$, and wave-vector 
$\vec{k}_L = k_L \vec{e}_z$
({\em inside} the BEC), and coherently excited atoms of density 
$\rho_e$ and of momentum $\hbar \vec{k}_L$. Inserting these expressions
into the Heisenberg equations of motions leads to a set of algebraical 
conditions which fix the chemical potential $\hbar\mu$,
the laser wavenumber $k_L$, and 
the density of coherently excited atoms $\rho_e$ which we assume to be 
smaller than $\rho_g$. 
If we neglect the kinetic energy the density of excited
atoms is given by $\sqrt{\rho_e} = \sqrt{\rho_g} \Omega_L/(\mu + \Delta_L)$.
The wavenumber $k_L$ is fixed by $c k_L = \omega_L/2 + \sqrt{(\omega_L/2)^2- 
\omega_{\mbox{{\scriptsize res}}}^2 \nu_g /(\mu+\Delta_L)}$,
where $\nu_g$ is defined in Eq.~(\ref{nug}). Note that $k_L$
generally is different from the free-space value $\omega_L/c$. 
For $\Delta_L \leq 0$ ($\Delta_L \geq 0$) the chemical potential is 
given by $\mu = -\Delta_L/2 \pm\sqrt{(\Delta_L/2)^2 +\Omega_L^2}$ 
which implies $\mu +\Delta_L >0$ ($\mu+\Delta_L<0$), respectively. Note
that for $\Delta_L>0$ the expression for $c k_L$ implies the
additional constraint $\mu+\Delta_L >4\nu_g$.
\subsection{Derivation of polariton eigenmodes}\label{polaritonEigenmodesRun}
Having found the macroscopic coherent solution we are in the position
to to derive the polariton eigenmodes 
Because of the delta distribution appearing in the macroscopic solution
(\ref{cohg}) the Hamiltonian (\ref{hpol}) reduces to a
sum of two-level systems so that its eigenmodes are quite easy to find.
They consist of polaritons with momentum $\hbar \vec{q}$ 
and frequency spectrum 
\begin{equation} 
  \omega_{\vec{q}, \pm} = {\Delta_q \over 2} \pm W_q\; ,
\end{equation}  
where we have defined 
$W_q := \sqrt{(\Delta_q/2)^2 + \nu_g \omega_{\mbox{{\scriptsize res}}} \sin^2 
\vartheta_{\vec{q}}}$ and $\Delta_q := c|\vec{q}| 
-\omega_{\mbox{{\scriptsize res}}}$. 
The coefficients of the polariton creation operator in Eq.~(\ref{linsup})
are given by
\begin{equation}  
  {\cal E}_{\vec{q},\pm} (\vec{k}) = 
    \frac{ \delta(\vec{q}-\vec{k}) }{\sqrt{2W_q}  }
    \frac{ \sqrt{\nu_g\omega_{\mbox{{\scriptsize res}}}} 
           \sin \vartheta_{\vec{q}}  }{\sqrt{ W_q \pm \Delta_q/2}  } 
    \quad ,  \quad 
  {\cal A}_{\vec{q},\pm} (\vec{k}) = \pm
     \frac{ \delta(\vec{q}-\vec{k}) }{\sqrt{2W_q}  }
  \sqrt{ W_q \pm \Delta_q/2 } 
\end{equation} 
so that we find for Eq.~(\ref{gqr})
\begin{equation} 
  g_{\vec{q},r}(\vec{k}) = (2\pi)^{3/2} \zeta(\vec{q}) \sqrt{\rho_e}
  \delta(\vec{k}+\vec{q}-\vec{k}_L) \frac{ \omega_{\vec{q},r} }{
   \sqrt{\omega_{\vec{q},r}^2  + \nu_g \omega_{\mbox{{\scriptsize res}}} \sin^2 \vartheta_{\vec{q}} }}
\label{gqrrun} \end{equation} 

The polariton spectrum $\omega_{\pm,\vec{q}}$  clearly
exhibits an avoided crossing around $\Delta_q =0$ of width
$\sqrt{\nu_g \omega_{\mbox{{\scriptsize res}}}} 
\sin \vartheta_{\vec{q}}$ (see Fig.~\ref{avoidedCrossing}). 
It also contains a small gap whose edge is reached 
in the limit $|\vec{q}| \rightarrow 0$ and $\infty$ \cite{shlyapnikov91}.
This gap will play no role for the SE rate since its edges are far away
from the resonance frequency. This is physically reasonable since
far away from resonance the two-level approximation for the atomic
internal structure ceases to be valid.
\subsection{Calculation of $I(s)$}\label{derivationOfIRun}
With the help of Eq.~(\ref{gqrrun}) the integral
$I(s)$ of Eq.~(\ref{integral}) can be written in the form
\begin{equation} 
  I = \frac{V \nu_e \omega_{\mbox{{\scriptsize res}}}^2}{(2\pi)^3} \int d^3k
    \frac{\sin^2 \vartheta_{\vec{k}}}{
    \omega_{\vec{k}} \left ( s-i\omega_L + i \omega_{\vec{k}}  +
    \frac{\nu_g \omega_{\mbox{{\scriptsize res}}} \sin^2 \vartheta_{\vec{k}} }{s-i\Delta_L}
    \right ) } \; ,
\label{Ihomog} \end{equation}
where $V$ denotes the quantization volume. 
This integral agrees with the one found in absence of a BEC 
(which describes SE in free space) by
setting $\nu_g = \Delta_L =0$. We denote this free space integral by
$I_0 :=I(\nu_g=\Delta_L=0)$. 
Both integrals are linearly divergent and can be
treated in the way pointed out by Bethe (see, e.g., Ref.~\cite{milonni}),
i.e., we renormalize the integrals by subtracting the free-electron 
contribution,
\begin{equation} 
  I^{\mbox{{\scriptsize Ren}}} := I -
  \frac{V \nu_e \omega_{\mbox{{\scriptsize res}}}^2}{(2\pi)^3 i} \int d^3k
    \frac{\sin^2 \vartheta_{\vec{k}}}{
    \omega^2_{\vec{k}} }
\end{equation} 
and $I_0^{\mbox{{\scriptsize Ren}}} = I^{\mbox{{\scriptsize Ren}}}(\nu_g=
\Delta_L=0)$. These renormalized integrals are only logarithmically
divergent. 

At this point it is customary in the calculation of the free-space
SE to perform the Wigner-Weisskopf approximation by
neglecting the dependence of  $I_0^{\mbox{{\scriptsize Ren}}}(s)$
on $s$. In the presence of a band gap this is inappropriate
due to the strong variation of the mode density around the gap
\cite{john94,kofman94}. Nevertheless, we can perform a generalized 
Wigner-Weisskopf approximation in the following way.
We expect that the typical 
timescale on which SE happens is much larger than the 
optical cycle timescale $1/\omega_{\mbox{{\scriptsize res}}}$. 
From the definition of the inverse Laplace transform (\ref{inverseLapl})
it is clear that the variable $s$ plays more or less the role of
a Fourier-transformed time. We thus expect that only values of $s$ much
smaller than $\omega_{\mbox{{\scriptsize res}}}$ 
contribute significantly to the SE.
This implies that we can neglect (the imaginary part of)
$s$ wherever it appears together with
$\omega_{\mbox{{\scriptsize res}}}$ or $\omega_L$. 
Thus, we are allowed to set $s-i\omega_L
\approx -i \omega_L$ in the denominator of $I(s)$ while retaining the
term depending on $s-i \Delta_L$.

In the case of $I_0^{\mbox{{\scriptsize Ren}}}$ this procedure
immediately reproduces the Wigner-Weisskopf result
$I_0^{\mbox{{\scriptsize Ren}}} \approx N_e \{ (
\gamma_{\mbox{{\scriptsize vac}}}/2) + i 
\Delta_{\mbox{{\scriptsize Lamb}}}^{\mbox{{\scriptsize 2-lev}}} \}$,
where  $\gamma_{\mbox{{\scriptsize vac}}}$ and $N_e$ are defined in
Sec.~\ref{resultsRun}.
To fix  $\Delta_{\mbox{{\scriptsize Lamb}}}^{\mbox{{\scriptsize 2-lev}}} $
we follow the theory of Bethe (see, e.g., \cite{milonni})
and introduce a cut-off frequency of $m_e c^2/\hbar$ in 
$I_0^{\mbox{{\scriptsize Ren}}}$, where $m_e$ is the electron's mass. 
Calculating the principal value of
the integral then leads to $\Delta_{\mbox{{\scriptsize Lamb}}}^
{\mbox{{\scriptsize 2-lev}}} \approx 2 \gamma_{\mbox{{\scriptsize vac}}}$.
In contrast to free space the SE rate in a BEC depends on
$\Delta_{\mbox{{\scriptsize Lamb}}}^{\mbox{{\scriptsize 2-lev}}} $
since such a radiative frequency correction shifts the center of the avoided 
crossing (or of a band gap \cite{kofman94}).

It remains to calculate a renormalized expression of the integral
$I^{\mbox{{\scriptsize Ren}}}$ in the presence of a BEC. Fortunately,
this task reduces to integrals proportional to 
$I_0^{\mbox{{\scriptsize Ren}}}$ and a couple of convergent integrals
and leads to Eq.~(\ref{iren})
\section{BEC in a standing wave laser} \label{standwavAppendix}
\subsection{Derivation of the macroscopic coherent 
solution} \label{classicalSolutionStand}
Since in a running laser wave the BEC density is not periodic no PBGs can be 
formed. It is therefore of interest to study a BEC interacting with a standing
laser wave so that the formation of photonic, or rather
polaritonic band gaps \cite{marzlin98b}, is possible. 
The coherent laser field describing a standing wave is given by 
\begin{equation} 
  a_\sigma^{\mbox{{\scriptsize coh}}}(\vec{k}) = 
  a_1^{\mbox{{\scriptsize coh}}} \delta_{\sigma,
  \sigma_0} \{ \delta(\vec{k}-\vec{k}_L) + \delta(\vec{k}+\vec{k}_L)\}
  \exp[-i \omega_L t] \; .
\label{classem} \end{equation} 
We assume that the amplitude $a_1^{\mbox{{\scriptsize coh}}}$ is real 
and that the polarization
$\sigma_0$ of the laser beam is parallel to the dipole moment $\vec{d}$
of the atoms. 

Since the laser field provides a periodic potential for the atoms it is 
reasonable to assume that the macroscopic atomic fields are periodic, too
(at least for the ground-state of the system). One also can make the ansatz
that $\Psi_g^{\mbox{{\scriptsize coh}}}$ has period $2 k_L$ so 
that the coherent solutions can be written as
\begin{eqnarray} 
  \Psi_g^{\mbox{{\scriptsize coh}}}(\vec{k}) &=& e^{-i \mu t} 
    \sum_n \delta (\vec{k} - 2n \vec{k}_L)
    \Psi_{g,2n}^{\mbox{{\scriptsize coh}}} \label{becg} \\
  \Psi_e^{\mbox{{\scriptsize coh}}}(\vec{k}) &=& e^{-i (\mu+\omega_L) t} 
   \sum_n \delta (\vec{k} - (2n+1) \vec{k}_L) 
   \Psi_{e,2n+1}^{\mbox{{\scriptsize coh}}} \; .
\label{bece} \end{eqnarray} 
Inserting this into Eqs.~(\ref{edgl}) to (\ref{adgl}) leads to the
matrix equations
\begin{eqnarray} 
   (\omega_L - c k_L) \Omega &=& {\zeta^2_{\sigma_0}(\vec{k}_L)
       \over \hbar^2} \sum_n \Psi_{g,2n}^{\mbox{{\scriptsize coh}}}
       \Psi_{e,2n+1}^{\mbox{{\scriptsize coh}}} \label{classa} \\
   \left (\Delta_L + \mu - (2n+1)^2 \frac{\hbar \vec{k}_L^2 }{2M}  \right ) 
       \Psi_{e,n+1}^{\mbox{{\scriptsize coh}}} &=& \Omega 
       \{ \Psi_{g,2n}^{\mbox{{\scriptsize coh}}} + 
          \Psi_{g,2n+2}^{\mbox{{\scriptsize coh}}} \} \label{classe} \\
   \left ( \mu - (2n)^2  \frac{\hbar \vec{k}_L^2 }{2M}  \right ) 
   \Psi_{g,2n}^{\mbox{{\scriptsize coh}}} &=&
   \Omega \{ \Psi_{e,2n-1}^{\mbox{{\scriptsize coh}}} + 
             \Psi_{e,2n+1}^{\mbox{{\scriptsize coh}}} \} \; . 
\label{classg} \end{eqnarray} 
We have assumed that $\zeta_{\sigma}(\vec{k})$ is real and does not
depend on the sign of $\vec{k}$ and introduced the real Rabi frequency
$\Omega := a_1^{\mbox{{\scriptsize coh}}} \zeta_{\sigma_0}(\vec{k}_L)/\hbar$.
For consistency with the assumption that $a_1^{\mbox{{\scriptsize coh}}}$ 
is a real quantity the
coefficients $\Psi_{g,n}^{\mbox{{\scriptsize coh}}}$ and 
$\Psi_{e,n}^{\mbox{{\scriptsize coh}}}$ must be real, too.

The system (\ref{classa})-(\ref{classg}) of algebraic equations can easily be
solved numerically. To do so we assume that the
Rabi frequency  of the laser
beam is a given quantity. For a given value of $k_L$ the
two equations (\ref{classe}) and (\ref{classg}) then just describe the
well-known problem of a two-level atom moving in a standing laser wave.
This is a simple system of linear equations and can be solved 
in a standard manner. The solution then can be inserted into 
Eq.~(\ref{classa}) which then, because $\omega_L$ and 
$a_1^{\mbox{{\scriptsize coh}}}$ are fixed, 
determines the value of $k_L$. We then have reinserted the
new value for  $k_L$ into the system (\ref{classa})-(\ref{classg})
and iterated the procedure until $k_L$ did not change significantly
anymore.
\subsection{Derivation of polariton eigenmodes}\label{polaritonEigenmodesStand}
In this case the periodic structure of the macroscopic solution
(\ref{becg}) leads to a more complicated structure
of the eigenmodes of $ H_{\mbox{{\scriptsize pol}}}$ than in the
case of a traveling laser wave.
To find these modes we make in Eq.~(\ref{linsup}) the ansatz
${\cal E}_{\vec{q},r} (\vec{k}) = \sum_{m\in {\mathbf Z}} 
{\cal E}_m(\vec{q},r) \delta(\vec{k}-\vec{q}-2m \vec{k}_L)$ and 
correspondingly for ${\cal A}_{\vec{q},r} (\vec{k})$. Now
$\vec{q}$ denotes the quasi-momentum of the polariton.
For a single standing laser wave along the z-axis
$q_z$ is confined to $[-k_L, k_L]$ whereas $q_x,q_y$ 
represent the real momentum of the polariton perpendicular to the laser beam.
The index $r$ is a collective notation for discrete quantum numbers
which include an internal quantum number taking
two values (since two quantum fields $\delta a$ and $\delta \Psi_e$ are
involved) and  the band index. This ansatz leads to
\begin{equation} 
  g_{\vec{q},r}(\vec{k}) := \sum_m {\cal A}_m(\vec{q},r) \zeta(\vec{q}+2m
  \vec{k}_L) \Psi_e^{\mbox{{\scriptsize coh}}}
  (\vec{k}+\vec{q}+2m \vec{k}_L) \; .
\label{gqrStand} \end{equation} 
and results in the eigenvalue equations
\begin{eqnarray} 
  E(\vec{q},r) {\cal E}_m(\vec{q},r) &=&-\hbar\Delta_L{\cal E}_m(\vec{q},r) + 
  \sum_n \Psi_{g,2n}^{\mbox{{\scriptsize coh}}} \zeta(\vec{q}+2(m-n)\vec{k}_L)
     {\cal A}_{m-n}(\vec{q},r)
  \\ 
  E(\vec{q},r) {\cal A}_m(\vec{q},r) &=& \hbar(c|\vec{q}+2m \vec{k}_L|
    -\omega_L) {\cal A}_m(\vec{q},r) +
  \sum_n \Psi_{g,2n}^{\mbox{{\scriptsize coh}}}\zeta(\vec{q}+2m\vec{k}_L) 
   {\cal E}_{m+n}(\vec{q},r)  \; .
\end{eqnarray} 
These equations can be substantially simplified by noting that the
frequency difference $\hbar(c|\vec{q}+2m \vec{k}_L|-\omega_L) $ is huge
compared to all other energy scales involved unless $m=0,\pm 1$ and
$\vec{q}$ is close to $\pm \vec{k}_L$. We thus can approximate the 
photon-part of all modes with $|m|>1$ as free photons and need only to
retain the coefficients ${\cal A}_0(\vec{q},r)$ and ${\cal A}_{-1}(\vec{q},r)$
for $q_z\in [0,k_L]$ and the coefficients ${\cal A}_0(\vec{q},r)$ and
${\cal A}_{1}(\vec{q},r)$ for $q_z\in [-k_L,0]$, respectively. We will 
focus here on the case  $q_z\in [0,k_L]$ since the second case can be
treated analogously.

To solve the resulting equations we introduce the two quantities
$F_0 := \sum_n \Psi_{g,2n}^{\mbox{{\scriptsize coh}}} 
{\cal E}_{n}(\vec{q},r)$ and $F_{-1} := \sum_n 
\Psi_{g,2n}^{\mbox{{\scriptsize coh}}} {\cal E}_{n-1}(\vec{q},r)$ 
and make the approximation $\zeta(\vec{q}) \approx \zeta (\vec{q}-2\vec{k}_L)
\approx \zeta(\vec{k}_L)$ so that the problem is reduced to the simple
matrix eigenvalue equation
\begin{equation} 
  ( E(\vec{q},r)+\hbar \Delta_L) \left ( \begin{array}{c}
    F_0 \\ F_{-1} \\ {\cal A}_0 \\ {\cal A}_{-1}  \end{array} \right ) =
    \left ( \begin{array}{cccc}
             0 & 0 & \zeta(\vec{k}_L) (2\pi)^3 \bar{\rho}_g   &  
                     \zeta(\vec{k}_L) (2\pi)^3\tilde{\rho}_g    \\
             0 & 0 & \zeta(\vec{k}_L) (2\pi)^3\tilde{\rho}_g  &  
                     \zeta(\vec{k}_L) (2\pi)^3 \bar{\rho}_g \\
              \zeta(\vec{k}_L) & 0 &\hbar \Delta_0 & 0  \\
             0 &  \zeta(\vec{k}_L) & 0 & \hbar\Delta_{-1}           
    \end{array} \right ) \left ( \begin{array}{c}
    F_0 \\ F_{-1} \\ {\cal A}_0 \\ {\cal A}_{-1}  \end{array} \right ) 
\label{44matrix} \end{equation} 
Here we have introduced 
\begin{eqnarray} 
  \Delta_0 &:=& c|\vec{q}|-\omega_{\mbox{{\scriptsize res}}} \\
  \Delta_{-1} &:=& c|\vec{q}-2\vec{k}_L|-\omega_{\mbox{{\scriptsize res}}}
\end{eqnarray} 
and the quantities $\bar{\rho}_g$ and $\tilde{\rho}_g$ which are
defined in Eqs.~(\ref{barrhog}) and (\ref{tilderhog}), respectively.

The problem of finding the eigenvalues and eigenvectors of a $4\times 4$
matrix is a basic one. The eigenvalues $\omega_{\vec{q},r} :=
(E(\vec{q},r)+\hbar \Delta_L)/\hbar$  fulfill the relation
$ P_{\mbox{{\scriptsize ch}}}(\omega_{\vec{q},r})=0$, where
\begin{equation}   
  P_{\mbox{{\scriptsize ch}}}(z) = (z^2-z \Delta_{-1} -\bar{\nu}_g
       \omega_{\mbox{{\scriptsize res}}} )
     (z^2-z \Delta_{0} -\bar{\nu}_g \omega_{\mbox{{\scriptsize res}}} ) -\tilde{\nu}_g^2 \omega_{\mbox{{\scriptsize res}}}^2
\label{pch} \end{equation} 
is the characteristic polynomial of the matrix and the frequencies
$ \bar{\nu}_g$ and $ \tilde{\nu}_g$ are defined in Eqs.~(\ref{barnug})
and (\ref{tildenug}).

\begin{figure}[t]
\epsfxsize=10cm
\hspace{1cm}
\epsffile{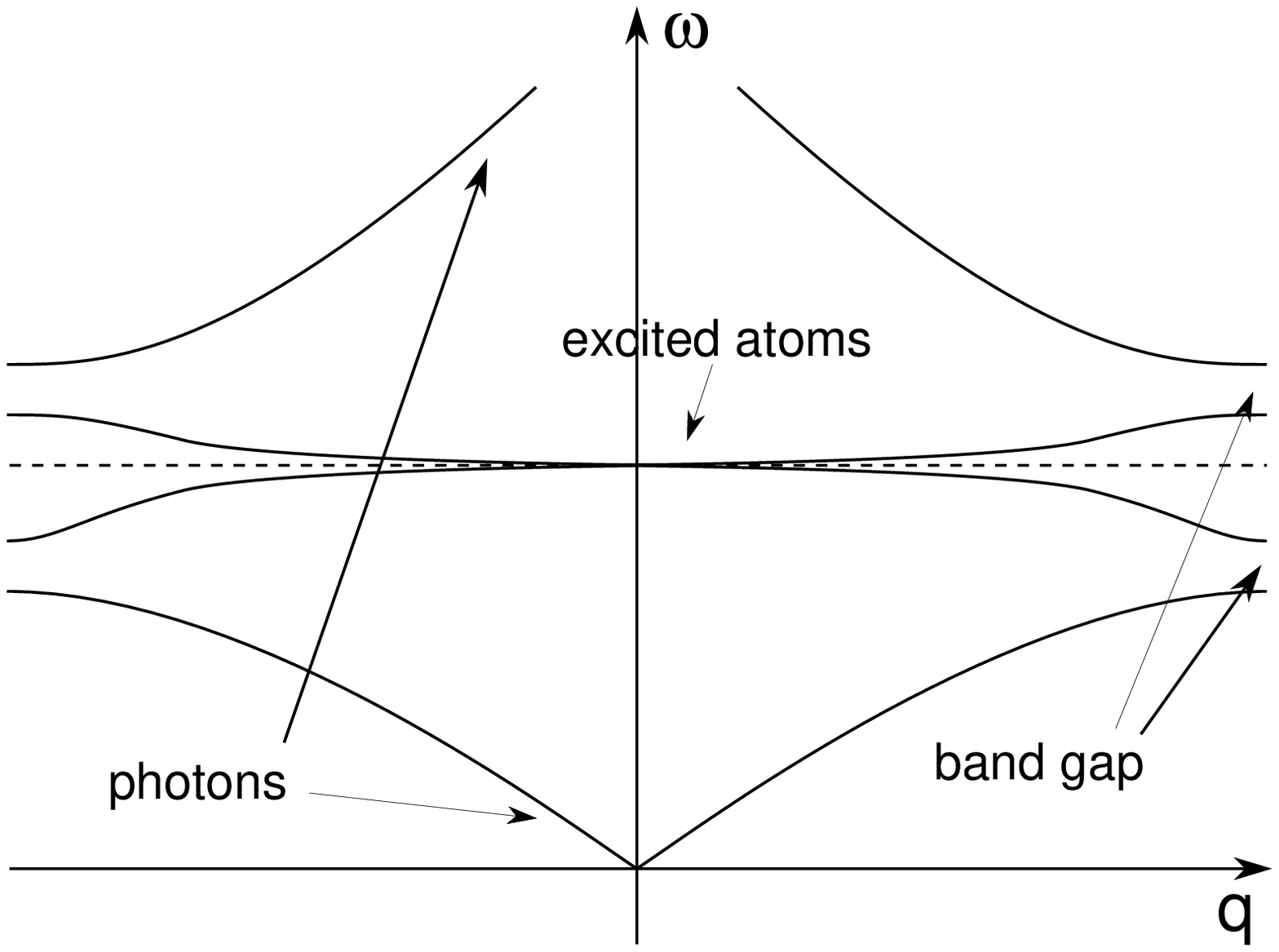}\vspace{-5mm}
\caption{\label{bandGap} A schematic drawing of the polariton spectrum for
a BEC in a 1D optical lattice. Because the BEC is periodic
band gaps do appear.}
\end{figure}

 Due to the periodicity of the ground-state BEC 
$\omega_{\vec{q},r}$ as a function of the quasi-momentum $\vec{q}$ exhibits
the phenomenon of band gaps \cite{marzlin98b}, see Fig.~\ref{bandGap}.
Though closed expressions for the eigenvalues 
$\omega_{\vec{q},r}$ do exist they are rather cumbersome and not of much 
use for our problem.

We instead will use a theorem
on the eigenvalues to derive the physical quantities of interest. To do
so we assume that we already know the eigenvalues. For a given
eigenvalue $\omega_{\vec{q},r}$ it is easy to solve for the eigenvectors.
For the relevant components we find
\begin{eqnarray} 
  {\cal A}_0(\vec{q},r) &=& \frac{   \omega_{\vec{q},r} 
   \sqrt{\omega_{\vec{q},r}^2- \omega_{\vec{q},r} \Delta_{-1}-\bar{\nu}_g
   \omega_{\mbox{{\scriptsize res}}}}
  }{ \sqrt{ 2\tilde{\nu}_g^2 \omega_{\mbox{{\scriptsize res}}}^2 + 
   (\omega_{\vec{q},r}^2 +\bar{\nu}_g \omega_{\mbox{{\scriptsize res}}})
   (2\omega_{\vec{q},r}^2 -\omega_{\vec{q},r}(\Delta_0+\Delta_{-1})
   -2\bar{\nu}_g \omega_{\mbox{{\scriptsize res}}} )  
  }}
  \label{beta0} \\
  {\cal A}_{-1}(\vec{q},r) &=& \frac{   \omega_{\vec{q},r} \tilde{\nu}_g 
   \omega_{\mbox{{\scriptsize res}}} /
   \sqrt{\omega_{\vec{q},r}^2- \omega_{\vec{q},r} \Delta_{-1}-
     \bar{\nu}_g \omega_{\mbox{{\scriptsize res}}}}
  }{
   \sqrt{ 2\tilde{\nu}_g^2 \omega_{\mbox{{\scriptsize res}}}^2 + 
   (\omega_{\vec{q},r}^2 +\bar{\nu}_g \omega_{\mbox{{\scriptsize res}}})
   (2\omega_{\vec{q},r}^2 -\omega_{\vec{q},r}(\Delta_0+\Delta_{-1})
   -2\bar{\nu}_g \omega_{\mbox{{\scriptsize res}}} )  
  }}
   \; .
\label{beta1} \end{eqnarray} 
The normalization has been done by requiring the particle number
(\ref{npol}) to be pseudo-normalized, i.e.,
\begin{equation} 
  \sum_m \left \{ |{\cal E}_m(\vec{q},r)|^2 +  |{\cal A}_m(\vec{q},r)|^2
  \right \} =1 \; .
\end{equation} 
The analogous equation for $q_z \in [-k_L,0]$ is obtained if
${\cal A}_{-1}(\vec{q},r)$ is replaced by ${\cal A}_{1}(\vec{q},r)$ and
$ \Delta_{-1}$ by 
$ \Delta_{1} := c|\vec{q}+2 \vec{k}_L| -\omega_{\mbox{{\scriptsize res}}}$.
\subsection{Calculation of $I(s)$}\label{derivationOfIStand}
To calculate the integral $I(s)$ of Eq.~(\ref{integral})
we first need to calculate the integral over $\vec{k}$
in the nominator which is an easy task. With Eq.~(\ref{gqrStand}) we find
for $q_z\in [0,k_L]$ 
\begin{equation} 
  \int d^3k |g_{\vec{q},r}(\vec{k})|^2 = \bar{N}_e \{ 
  {\cal A}_0^2 (\vec{q},r) \zeta^2(\vec{q}) +
  {\cal A}_{-1}^2 (\vec{q},r) \zeta^2(\vec{q}-2 \vec{k}_L) \}  +
  2 \tilde{N}_e {\cal A}_{-1}(\vec{q},r) {\cal A}_0(\vec{q},r)
   \zeta(\vec{q}) \zeta(\vec{q}-2 \vec{k}_L)
\label{betasum} \end{equation} 
if $r$ is in the lowest two energy bands. The numbers $\bar{N}_e$
and $\tilde{N}_e$ are defined in Eqs.~(\ref{barNe}) and (\ref{tildeNe}),
respectively.
For $q_z\in [-k_L,0]$ and the two lowest energy bands
one has to replace ${\cal A}_{-1}(\vec{q},r)$
by  ${\cal A}_{1}(\vec{q},r)$ and $\zeta(\vec{q}-2 \vec{k}_L)$ by
$\zeta(\vec{q}+2 \vec{k}_L)$.
For all higher bands, which
according to our approximation just describe free photons, 
the integral (\ref{betasum}) has the simple value $\bar{N}_e \zeta^2(\vec{q})$.

From Eq.~(\ref{betasum}) it becomes clear that
to calculate $I$ of Eq.~(\ref{integral}) we have to find closed expressions
for terms like $\sum_r {\cal A}_{-1}(\vec{q},r) {\cal A}_{0}(\vec{q},r) /
(z_s - \omega_{\vec{q},r})$. For the lowest two energy bands,
the sum over $r$ now runs over the four eigenvectors of the matrix 
(\ref{44matrix}). It would be extremely tedious if not 
practically impossible
to derive these sums by simply inserting the complicated closed expressions
for $ \omega_{\vec{q},r}$ into them. Instead, we start with the observation
that the polynomial appearing in the denominator of Eqs.~(\ref{beta0})
and (\ref{beta1}) can be written as
\begin{equation} 
  2\tilde{\nu}_g^2 \omega_{\mbox{{\scriptsize res}}}^2 + (\omega_{\vec{q},r}^2 +\bar{\nu}_g\omega_{\mbox{{\scriptsize res}}})
   (2\omega_{\vec{q},r}^2 -\omega_{\vec{q},r}(\Delta_0+\Delta_{-1})
   -2\bar{\nu}_g\omega_{\mbox{{\scriptsize res}}} )   =   
   \omega_{\vec{q},r} P_{\mbox{{\scriptsize ch}}}^\prime (
   \omega_{\vec{q},r}) \; ,
\end{equation} 
where $P_{\mbox{{\scriptsize ch}}}^\prime(z)$ denotes the derivative
of the characteristic polynomial (\ref{pch}). This enables us to write the
sum under consideration in the form
\begin{equation} 
  \sum_r \frac{{\cal A}_{-1}(\vec{q},r) {\cal A}_{0}(\vec{q},r)   }{
  (z_s - \omega_{\vec{q},r})} =
  \tilde{\nu}_g\omega_{\mbox{{\scriptsize res}}} \sum_r \frac{\omega_{\vec{q},r}          }{
   P_{\mbox{{\scriptsize ch}}}^\prime (\omega_{\vec{q},r}) 
   (z_s - \omega_{\vec{q},r})  } \; .
\end{equation} 
This can be further simplified by noting that the characteristic polynomial
can also be written in the form
$ P_{\mbox{{\scriptsize ch}}}(z) = \prod_r (z- \omega_{\vec{q},r})$ so that
we have $P_{\mbox{{\scriptsize ch}}}^\prime (\omega_{\vec{q},r}) =
\prod_{r^\prime \neq r} (\omega_{\vec{q},r} -\omega_{\vec{q},r^\prime })$.
Using this expression it is straightforward if still tedious to find
\begin{equation} 
  \sum_r \frac{{\cal A}_{-1}(\vec{q},r) {\cal A}_{0}(\vec{q},r)   }{
  (z_s - \omega_{\vec{q},r})} =
  \frac{z_s \tilde{\nu}_g\omega_{\mbox{{\scriptsize res}}}}{P_{\mbox{{\scriptsize ch}}}(z_s) } \; .
\label{sum01}\end{equation} 
We thus have been able to calculate this sum without explicit knowledge
of the eigenvalues of Eq.~(\ref{44matrix}). The remaining sums which
result from the insertion of Eq.~(\ref{betasum}) into Eq.~(\ref{integral})
can be treated in a similar way and are given by
\begin{eqnarray} 
  \sum_r \frac{{\cal A}_{0}^2(\vec{q},r)  }{
  (z_s - \omega_{\vec{q},r})} &=& z_s 
  \frac{z_s^2- z_s \Delta_{-1}-\bar{\nu}_g\omega_{\mbox{{\scriptsize res}}} 
  }{  P_{\mbox{{\scriptsize ch}}}(z_s)  }
  \label{sum00} \\
  \sum_r \frac{{\cal A}_{-1}^2(\vec{q},r)  }{
  (z_s - \omega_{\vec{q},r})} &=& z_s
  \frac{z_s^2- z_s \Delta_{0}-\bar{\nu}_g\omega_{\mbox{{\scriptsize res}}} 
  }{  P_{\mbox{{\scriptsize ch}}}(z_s)  }
\label{sum11} \end{eqnarray} 
for the lowest two energy bands and $q_z \in [0,k_L]$.
Again the corresponding expressions for  $q_z \in [-k_L,0]$ are obtained
by replacing ${\cal A}_{-1}(\vec{q},s)$  by ${\cal A}_{1}(\vec{q},s)$ and
$ \Delta_{-1}$ by $ \Delta_{1}$.

Taking everything together the use of Eq.~(\ref{betasum})
and the three sums (\ref{sum01}), (\ref{sum00}), and (\ref{sum11})
allows us to bring the integral (\ref{integral}) into the form
\begin{equation} 
\begin{array}{l} \displaystyle
   I= 
\\[5mm]  \displaystyle 
 \frac{\bar{N}_e}{i\hbar^2} \int_{|q_z|>2k_L}dq_z 
    \int_{-\infty}^\infty dq_x dq_y \frac{\zeta(\vec{q})^2 }{
    z_s-c|\vec{q}| +\omega_{\mbox{{\scriptsize res}}}} +
\\[5mm] \displaystyle  
  \frac{z_s}{i\hbar^2} \int_{V_-}
  \frac{ \bar{N}_e\zeta(\vec{q})^2 [z_s^2-z_s \Delta_{1}-\bar{\nu}_g
        \omega_{\mbox{{\scriptsize res}}}]
     +  \bar{N}_e\zeta(\vec{q}+2\vec{k}_L)^2 
        [z_s^2-z_s \Delta_{0}-\bar{\nu}_g\omega_{\mbox{{\scriptsize res}}}]
     +2\tilde{N}_e \tilde{\nu}_g\omega_{\mbox{{\scriptsize res}}}
        \zeta(\vec{q})\zeta(\vec{q}+2\vec{k}_L)
  }{
     z_s^4-z_s^3(\Delta_0+\Delta_{1})+z_s^2(\Delta_0\Delta_{1}-2
     \bar{\nu}_g\omega_{\mbox{{\scriptsize res}}}) +z_s \bar{\nu}_g
     \omega_{\mbox{{\scriptsize res}}}(\Delta_0+\Delta_{1}) +
     \bar{\nu}_g^2\omega_{\mbox{{\scriptsize res}}}^2 -\tilde{\nu}_g^2
     \omega_{\mbox{{\scriptsize res}}}^2 
  } + 
\\[5mm] \displaystyle  
  \frac{z_s}{i\hbar^2} \int_{V_+}d^3q
  \frac{ \bar{N}_e\zeta(\vec{q})^2 [z_s^2-z_s \Delta_{-1}-\bar{\nu}_g
        \omega_{\mbox{{\scriptsize res}}}]
     +  \bar{N}_e\zeta(\vec{q}-2\vec{k}_L)^2 
        [z_s^2-z_s \Delta_{0}-\bar{\nu}_g\omega_{\mbox{{\scriptsize res}}}]
     +2\tilde{N}_e \tilde{\nu}_g\omega_{\mbox{{\scriptsize res}}}
       \zeta(\vec{q})\zeta(\vec{q}-2\vec{k}_L)
  }{
     z_s^4-z_s^3(\Delta_0+\Delta_{-1})+z_s^2(\Delta_0\Delta_{-1}-2
     \bar{\nu}_g\omega_{\mbox{{\scriptsize res}}}) +z_s \bar{\nu}_g
     \omega_{\mbox{{\scriptsize res}}}(\Delta_0+\Delta_{-1}) +
     \bar{\nu}_g^2\omega_{\mbox{{\scriptsize res}}}^2 -\tilde{\nu}_g^2
     \omega_{\mbox{{\scriptsize res}}}^2 
  } 
\end{array}
\label{integ1} \end{equation} 
where the areas of integration $V_\pm$ are given by
$q_x,q_y \in (-\infty,\infty)$ and $q_z\in [0,k_L]$ for $V_+$ as well as
$q_z \in [-k_L,0]$ for $V_-$. The first integral in Eq.~(\ref{integ1})
represents the contribution from the higher polaritonic energy bands 
where the polaritons can be considered as free photons. Apart from
the restriction $|q_z|>2k_L$, which essentially means that the photons
are far off-resonant, it has the same form as the integral
\begin{equation} 
  I_0 = \frac{\bar{N}_e}{i\hbar^2} \int_{-\infty}^\infty dq_x dq_y dq_z
     \frac{\zeta(\vec{q})^2 }{
    z_s-c|\vec{q}| +\omega_{\mbox{{\scriptsize res}}}}
\label{i0} \end{equation} 
which appears in the calculation of the free-space SE rate.
The second contribution arises from the two lowest energy bands for
negative quasi-momentum $0>q_z>-k_L$, and the third integral represents the
corresponding contribution for positive quasi momentum $q_z$. It is not
hard to see that Eq.~(\ref{integ1}) reduces to the free integral (\ref{i0})
in absence of a ground-state BEC, i.e., for $\bar{\nu}_g=\tilde{\nu}_g=0$
and to Eq.~(\ref{Ihomog}) if
the BEC is homogeneous (for $\tilde{\nu}_g=0$ but $\bar{\nu}_g\neq0$).

We now return to the evaluation of Eq.~(\ref{integ1}). Shifting the
integration variable $q_z$ and exploiting the symmetries
$\zeta(-\vec{k})=\zeta(\vec{k})$ and $\Delta_1(-\vec{q}) =
\Delta_{-1}(\vec{q})$ allows us to combine the last two integrals
into a more convenient form. To simplify the process of renormalization
it is also advantageous to calculate $I-I_0$ instead of $I$ alone
since in this difference the divergence appearing in $I_0$ is canceled.
We then find
\begin{eqnarray} 
  I-I_0 &=& \frac{2i\bar{N}_e}{\hbar^2} \int_0^{2k_L}dq_z 
    \int_{-\infty}^\infty dq_x dq_y \frac{\zeta(\vec{q})^2 }{
    z_s-c|\vec{q}| +\omega_{\mbox{{\scriptsize res}}}} +
  \label{integ2} \\ & &
  \frac{2 z_s}{i\hbar^2} \int_0^{2k_L}dq_z \int dq_xdq_y
  \frac{ \bar{N}_e\zeta(\vec{q})^2 [z_s^2-z_s \Delta_{-1}-\bar{\nu}_g\omega_{\mbox{{\scriptsize res}}}]
     +2\tilde{N}_e \tilde{\nu}_g\omega_{\mbox{{\scriptsize res}}}\zeta(\vec{q})\zeta(\vec{q}+2\vec{k}_L)
  }{
    [ z_s^2 -z_s \Delta_0 - \bar{\nu}_g\omega_{\mbox{{\scriptsize res}}}]
    [ z_s^2 -z_s \Delta_{-1} - \bar{\nu}_g\omega_{\mbox{{\scriptsize res}}}] -
    \tilde{\nu}_g^2\omega_{\mbox{{\scriptsize res}}}^2 
  } 
 \nonumber 
\end{eqnarray} 
To calculate these expressions we have to make one further approximation
by neglecting the angular dependence in $\zeta(\vec{q}) =
\sin(\theta)\omega_{\mbox{{\scriptsize res}}} |\vec{d}| [\hbar/(2(2\pi)^3\varepsilon_0 c |\vec{q}|
]^{1/2}$, where $\theta$ is the angle between $\vec{q}$ and the atomic
dipole moment $\vec{d}$. Replacing $\sin(\theta)$ by $\sqrt{2/3}$
for all values of $\vec{q}$ leads to the correct result in absence of
a BEC and should produce qualitatively correct results for the case
under consideration. Doing this approximation in
the case of a homogeneous BEC 
just amounts in replacing a factor of $4/5$ by $2/3$ in the modifications
of the SE rate (\ref{homogExpan}). 
We remark that neglecting the dependence of $\zeta$ on $\sin(\theta)$
does only symmetrize the integrand in the $(q_x,q_y)$ plane. This does
not correspond to an isotropic band model because the asymmetry
between $q_z$ and $(q_x,q_y)$ still persists.

The calculation of the first integral of Eq.~(\ref{integ2}), which
roughly corresponds to the contribution of $-I_0$, is quickly done
and results in
\begin{equation} 
  \frac{2i\bar{N}_e}{\hbar^2} \int_0^{2k_L}dq_z \int_{-\infty}^\infty 
   dq_x dq_y \frac{\zeta(\vec{q})^2 }{ z_s-c|\vec{q}| +\omega_{\mbox{{\scriptsize res}}}} =
   \frac{\bar{N}_e \gamma_{\mbox{{\scriptsize vac}}}}{2\pi i} \left \{ 2 
   + 2 \ln \left (
   \frac{\Lambda}{\omega_{\mbox{{\scriptsize res}}}} \right ) + i \pi \mbox{sgn(Im} (z_s))
   \right \} \; .
\label{freeContrib}\end{equation} 
Here $\Lambda$ is a cut-off which
usually is taken to be $\Lambda = M_e c^2/\hbar$, where $M_e$ is
the electron's mass (see, e.g., Ref.~\cite{milonni}).

The dependence on the sign of Im($z_s$) originally comes from a
logarithm of the form $\ln (-(\omega_{\mbox{{\scriptsize res}}}+z_s)/\Lambda)$. This expression
can be reduced to the one presented in Eq.~(\ref{freeContrib})
by doing a generalized Wigner-Weisskopf approximation as it was introduced
above in the case of a running laser wave.
Though Im($z_s$) is also much smaller than any other
quantity in the above logarithm, it determines the sign of the
imaginary part of the logarithm's argument. Since the logarithm has
a branch cut along the negative real axis, this sign determines
on which side of the cut we are. 

It is also worth remarking that Eq.~(\ref{freeContrib}) is logarithmically
divergent with $\Lambda$ although we did not subtract the free-electron
part, a step which in the free-space calculation is done to remove
a linearly divergent contribution (see, e.g., Ref.~\cite{milonni}).
This is because the integration over $q_z$ does not extend to infinity,
thus reducing the degree of divergence by one. 

As has been already mentioned, Eq.~(\ref{freeContrib}) very roughly
corresponds to the negative of the free-space integral $I_0$. As
a consequence, its contribution will be mostly canceled after the
renormalization of $I$. This renormalization is easily done by noting that
$I-I_0 = I^{\mbox{{\scriptsize Ren}}} -I_0^{\mbox{{\scriptsize Ren}}}$
so that $I^{\mbox{{\scriptsize Ren}}} = (I-I_0) + 
I_0^{\mbox{{\scriptsize Ren}}}$, where the superscript ``Ren''
denotes the renormalized integrals and the renormalized free-space integral
is approximately given by
\begin{equation} 
  I_0^{\mbox{{\scriptsize Ren}}} = i \frac{\gamma_{\mbox{{\scriptsize vac}}} 
  \bar{N}_e}{2\pi}
  \left \{ \ln \left ( \frac{\Lambda}{\omega_{\mbox{{\scriptsize res}}}}\right ) + i \pi
  \mbox{sgn(Im}(z_s)) \right \} \; .
\end{equation} 
This allows us to derive from Eq.~(\ref{integ2}) the expression 
\begin{eqnarray} 
  I^{\mbox{{\scriptsize Ren}}} &=& 
     \frac{ \bar{N}_e \gamma_{\mbox{{\scriptsize vac}}} }{ 2\pi i } 
    \left  \{ 2 + \ln \left (\frac{\Lambda}{\omega_{\mbox{{\scriptsize res}}}} \right ) 
    \right \} +
  \label{mainresult} \\ & &
  \frac{ z_s \bar{N}_e \gamma_{\mbox{{\scriptsize vac}}} c^2}{ 4i \pi^2 
  \omega_{\mbox{{\scriptsize res}}} } \int_0^{2k_L} dq_z \int dq_x dq_y
  \frac{ [z_s^2 -z_s \Delta_{-1} - \bar{\nu}_g\omega_{\mbox{{\scriptsize res}}}]
  }{ |\vec{q}|
    \left \{
      [ z_s^2 -z_s \Delta_0 - \bar{\nu}_g\omega_{\mbox{{\scriptsize res}}}]
      [ z_s^2 -z_s \Delta_{-1} - \bar{\nu}_g\omega_{\mbox{{\scriptsize res}}}] -
      \tilde{\nu}_g^2 \omega_{\mbox{{\scriptsize res}}}^2 
    \right \}
  } +
  \nonumber \\ & &
  \frac{z_s\tilde{N}_e \tilde{\nu}_g  \gamma_{\mbox{{\scriptsize vac}}}c^3 }{
  2i\pi^2} \int_0^{2k_L}dq_z \int dq_xdq_y
  \frac{|\vec{q}|^{-1/4} |\vec{q}+2 \vec{k}_L|^{-1/4}
  }{
      [ z_s^2 -z_s \Delta_0 - \bar{\nu}_g\omega_{\mbox{{\scriptsize res}}}]
      [ z_s^2 -z_s \Delta_{-1} - \bar{\nu}_g\omega_{\mbox{{\scriptsize res}}}] -
      \tilde{\nu}_g^2\omega_{\mbox{{\scriptsize res}}}^2 
  } 
 \nonumber 
\end{eqnarray}
To reduce Eq.~(\ref{mainresult}) it is
convenient to introduce the scaled variables of integration
$u:=  q_z/(2 k_L)$ and $v := (q_x^2 + q_y^2)/(4 k_L^2)$
and the abbreviations (\ref{f0}) and (\ref{f1})
for the evaluation. Eq.~(\ref{mainresult}) then becomes
\begin{eqnarray} 
  I^{\mbox{{\scriptsize Ren}}} &=& 
     \frac{ \bar{N}_e \gamma_{\mbox{{\scriptsize vac}}} }{ 2\pi i } 
    \left  \{ 2 + \ln \left (\frac{\Lambda}{\omega_{\mbox{{\scriptsize res}}}} \right ) 
    \right \} +
  \label{integ3} \\ & &
  \frac{\gamma_{\mbox{{\scriptsize vac}}}\bar{N}_e}{2\pi i} 
  \int_0^1 du \int_0^\infty dv
  \frac{ (c k_L/\omega_{\mbox{{\scriptsize res}}})  [ f_0(z_s) - \sqrt{(u-1)^2+v} ]
  }{
  \sqrt{u^2+v}
  \{[f_0(z_s)-\sqrt{u^2+v}] [f_0(z_s) -\sqrt{(u-1)^2+v}] -f_1(z_s) \}
  } +
  \nonumber \\ & &
  \frac{\gamma_{\mbox{{\scriptsize vac}}}\tilde{N}_e}{4\pi i} 
  {\tilde{\nu}_g\over z_s} \int_0^1 du \int_0^\infty dv
  \frac{(u^2+v)^{-1/4} ((u-1)^2+v)^{-1/4}
  }{ 
    [f_0(z_s)-\sqrt{u^2+v}] [f_0(z_s) -\sqrt{(u-1)^2+v}] -f_1(z_s)
  }
 \nonumber 
\end{eqnarray} 
This expression can be further reduced by switching to the integration 
variable $v^\prime = \sqrt{(u-1)^2+v}$ and exchanging the sequence of
integration so that one first integrates over $u$. This allows us
to reduce the integral $I$ to a number of one-dimensional integrals
and leads us to our final analytical result (\ref{analyticalResult}).
\end{appendix}

\newpage

\begin{thebibliography}{99}
\bibitem{cavity} P. Goy, J.M. Raimond, M. Gross, and S. Haroche,
	Phys.~Rev.~Lett. {\bf 50}, 1903 (1983).
\bibitem{john94} S. John and T. Quang, Phys.~Rev.~A {\bf 50}, 1764 (1994). 
\bibitem{kofman94} A.G. Kofman, G. Kurizki, and B. Sherman, 
	J.~Mod.~Opt.~{\bf 41}, 353 (1994).
\bibitem{experimente} M. Anderson, J.R. Ensher, M.R. Matthews, 
	C.E. Wieman, and E.A. Cornell, Science {\bf 269}, 198 (1995);
	C.C. Bradley, C.A. Sackett, J.J. Tollet and R. Hulet, 
	Phys.~Rev.~Lett.~{\bf 75}, 1687 (1995);
	M.-O. Mewes, M.R. Andrews, N.J. van Druten, D.M. Kurn, 
	D.S. Durfee, C.G. Townsend and W. Ketterle, 
	Phys.~Rev.~Lett. {\bf 77}, 416 (1996).
\bibitem{shlyapnikov91} B.V. Svistunov and G.V. Shlyapnikov,
	Sov. Phys. JETP {\bf 71}, 71 (1990);
        H.D. Politzer, Phys. Rev. A {\bf 43}, 6444 (1991).
\bibitem{javanainen93} J. Javanainen, Phys. Rev. Lett. {\bf 72}, 2375 (1994).
\bibitem{cooper95} L. You, M. Lewenstein, R.J. Glauber, and J. Cooper,
	Phys. Rev. A {\bf 53}, 329 (1996).
\bibitem{hope97} J.J.~Hope and C.M.~Savage, Phys. Rev. A {\bf 54}, 3177 (1996).
\bibitem{savage97} C.M. Savage, J. Ruostekoski, and D.F. Walls,
	Phys. Rev. A {\bf 56}, 2046 (1997).
\bibitem{marzlin98b} K.-P. Marzlin and W. Zhang, Phys.~Rev.~A.
	{\bf 59}, 2982 (1999).
\bibitem{rzazewski92} K. Rz\c{a}zewski and W. Zakowicz, Journ. Phys. B 
	{\bf 25}, L319 (1992).
\bibitem{milonni} P.W. Milonni, {\em The quantum vacuum}, Academic Press,
	San Diego 1994.
\bibitem{john90} S. John and J. Wang, Phys.~Rev.~Lett. {\bf 64}, 2418 (1990).
\end{thebibliography}
\end{document}